\documentclass[%
preprint,
 amsmath,amssymb,
 aps, nofootinbib
]{revtex4-2}

\usepackage{graphicx}
\usepackage{dcolumn}
\usepackage{bm}
\usepackage{physics}
\usepackage{xcolor}
\usepackage{appendix}

\usepackage{caption}
\usepackage{subcaption}

\usepackage{bbold}
\begin{document}
\title{An Operational Quantum Field Theoretic Model for Gravitationally Induced Entanglement}

\author{Jackson Yant}%
\email{Jackson.R.Yant.Gr@Dartmouth.edu}

\affiliation{Department of Physics and Astronomy, Dartmouth College, Hanover, New Hampshire 03755, USA}

\author{Miles Blencowe}
\email{Miles.P.Blencowe@Dartmouth.edu}

\affiliation{Department of Physics and Astronomy, Dartmouth College, Hanover, New Hampshire 03755, USA}

\date{\today}%

\begin{abstract}
We develop a quantum field-theoretic model of gravitationally induced entanglement (GIE) between two massive objects in spatial superposition. The masses are described as excitations of a scalar field in an external harmonic potential, allowing for a well-defined notion of relativistic coherent states. Using linearized quantum gravity in the static limit, we derive an effective Hamiltonian that induces entanglement between the field modes occupied by the masses.

To probe this entanglement, we construct an observable from field operators that corresponds to the probability density of detecting the center of mass of one of these massive objects. Using this, we compute the fringe visibility in the overlap region and find that gravitationally induced entanglement leads to a decrease in visibility, consistent with previous nonrelativistic results. Additionally, we identify relativistic corrections that accelerate the decay of fringe visibility. These results provide a framework for studying weakly relativistic quantum field systems and their gravitational interactions in tabletop experiments.
\end{abstract}

\maketitle
\newpage
\section{\label{sec:introduction} Introduction}
Gravity and quantum mechanics form the two cornerstones of modern physics, yet unifying them into a consistent theory of quantum gravity remains one of the most enduring challenges in theoretical physics \cite{macias_incompatibility_2008,loll_quantum_2022,carlip_quantum_2001,kiefer_quantum_2006}. This difficulty arises in part from the weakness of gravity at small scales, which makes it extremely difficult to observe quantum gravitational effects directly. Without experimental guidance, theoretical progress has been slow, leaving the question of how quantum mechanics and gravity can coexist largely unresolved \cite{kiefer_conceptual_2013,oppenheim_is_2023}.

While the inherent weakness of gravitational effects at small scales has made experimental progress in quantum gravity slow, recent advances have offered new hope. The initial proposals from Bose et al. \cite{bose_spin_2017} and Marletto and Vedral \cite{marletto_gravitationally_2017} introduced gravitationally induced entanglement (GIE) as a potentially observable consequence of quantum gravitational effects, leading to the possibility of testing gravity's quantum nature in tabletop experiments. These groundbreaking ideas have inspired a broad spectrum of research, with some focusing on how to realize the experimental proposals \cite{yi_spatial_2022,carney_using_2021,cui_exponentially_2023,feng_amplification_2022,guff_optimal_2022,tilly_qudits_2021,fujita_inverted_2023,kaku_sudden_2025,miki_quantum_2023,schut_relaxation_2023,patrascu_graviton_2023}, others exploring the theoretical implications of GIE in understanding the nature of quantum gravity \cite{anastopoulos_gravity_2022,christodoulou_locally_2023,christodoulou_gravity_2023, bose_mechanism_2022, danielson_gravitationally_2022,ma_limits_2022,marletto_when_2018,martin-martinez_what_2023,rydving_gedanken_2021,di_biagio_relativistic_2023,carney_newton_2022}, and yet others pursuing entirely different ideas for testing quantum gravity that might not have emerged without the renewed interest in this field \cite{chen_quantum_2024,higgins_gravitationally_2024,kaku_gravitational_2024,tobar_detecting_2023,bose_massive_2023,howl_non-gaussianity_2021,maleki_complementarity-entanglement_2022,carney_quantum_2025,kryhin_distinguishable_2023}.

Given the promising implications of gravitationally induced entanglement, it is essential to deepen our understanding of this phenomenon to fully explore its potential within both quantum mechanics and general relativity. To this end, the present work contributes to the growing body of research by developing a model of gravitationally induced entanglement from first principles, within the foundational framework of quantum field theory. 

Our model considers two masses in a harmonic trap, each prepared in a superposition of coherent states. As the system evolves, gravitational interactions generate entanglement between them. The trap functions as an interferometer, periodically causing the coherent state components of each mass to overlap and produce an interference pattern in the probability of detection. This setup enables us to study entanglement signatures without requiring external beam splitters or mirrors, which would introduce singular potentials into a field-theoretic framework. A more detailed description of this configuration is provided in Section \ref{sec:set up}.

Unlike previous approaches that model the masses which become entangled as point particles, we treat them as states of a scalar field with an additional spatially dependent external potential. In this framework, the masses are represented as many-particle excitations of the scalar field. To construct the coherent states forming the components of the matter field, we analyze the action of the field's translation operator on the many-particle ground state. In the weakly relativistic limit, we find that the lowest-order corrections appear as a perturbation to the frequency, dependent on the amplitude of the coherent state. Additionally, the variance of the state exhibits oscillations that grow over time. A more detailed discussion of this derivation is provided in Section \ref{sec:scalar field}.

In order to describe the gravitational interaction of the masses we work with linearized quantum gravity as an effective field theory \cite{donoghue_effective_2012}. Though not a complete theory of quantum gravity, linearized gravity provides a well-established effective field theory suitable for exploring GIE in experimentally relevant regimes. To simplify the model, we focus on the static limit, where we neglect all but the energy density component of the stress-energy tensor. Within this approximation, we derive a representation of the time evolution operator, which allows the linearized gravitational field to induce a self-interaction of the matter field’s energy density, while the matter field itself shifts the vacuum of the gravitational field. This approach isolates the component that entangles the matter field with the gravitational field—an interaction that we argue can be neglected in this context. Tracing out the gravitational field’s state yields effective Hamiltonian dynamics that reveal the components responsible for entangling the matter field's degrees of freedom. Further details of this derivation are provided in Section \ref{sec:LQG}.

Using this effective Hamiltonian derived in the previous section, we approximate the time evolution of the system, treating the gravitational interaction as a perturbation to the harmonic trapping potential. We adapt a semiclassical approximation method developed by Heller \cite{heller_time-dependent_1975}, previously applied in our non-relativistic analysis \cite{yant_gravitationally_2023}, to determine the phase acquired by each coherent state component. This phase shift, induced by the gravitational interaction, leads to entanglement between the masses. We extend this approximation to all components of the initial state, resulting in an entangled final state.

We adopt an operational approach, evaluating entanglement through expectation values of field observables rather than relying on non-relativistic quantum mechanics (NRQM) for measurement. Specifically, we construct a field operator sensitive to the center-of-mass degree of freedom and use its expectation value to compute the probability density and fringe visibility.

The result of the fringe visibility calculation is then compared with that of a non-relativistic quantum mechanical model of the same system, where two particles in a harmonic trap interact through a Newtonian gravitational potential. In this comparison, we employ an updated version of the NRQM model from our previous work \cite{yant_gravitationally_2023}, which includes a more rigorous calculation of the phases acquired by each component. Our results show that, in the limit where the kinetic (or potential) energy of the coherent states is much smaller than the rest mass energy of the particles, the operational field-theoretic model approaches the NRQM model. We also find lowest order relativistic corrections which accelerate the reduction of fringe visibility. The details of this comparison are discussed in Section \ref{sec:QFT GIE}.

Finally, in the conclusion (Section \ref{sec:conclusion}), we summarize the significance of these results, focusing on how the operational field-theoretic approach deepens our understanding of gravitationally induced entanglement. We also discuss the application of the static limit of linearized quantum gravity and its potential to provide new insights into the dynamics of entangled states in quantum gravity. Additionally, we suggest possible directions for future research.

\section{\label{sec:set up}Conceptual Setup for Gravitationally Induced Entanglement in a Harmonic Trap}

To further explore gravitationally induced entanglement, it is essential to define a setup that allows for a field-theoretic description, where the preparation of states and measurement of observables can be articulated in an operational manner. In this section, we outline such a setup, presenting it in a model-independent way that can be applied both to the relativistic field-theoretic framework developed here and to the simpler non-relativistic quantum mechanical model from our previous work. This provides a foundation for the subsequent analysis and comparisons between different theoretical approaches. While this setup is not intended as a practical experimental proposal for detecting GIE, it serves as a controlled theoretical setting that naturally lends itself to a field-theoretic treatment, making it well-suited for our analysis.

\begin{figure}
    \centering
    \includegraphics[scale=0.26]{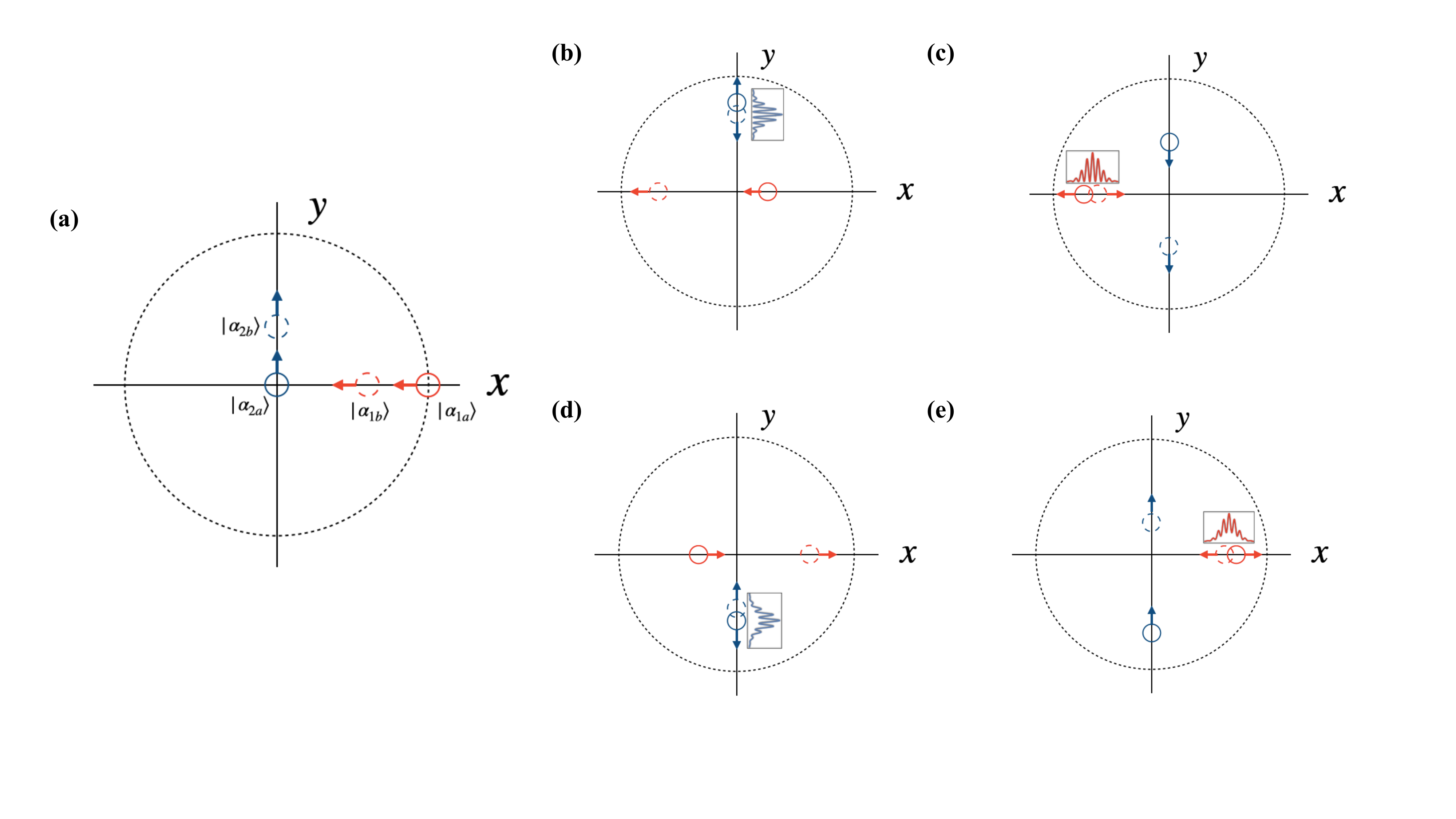}
    \caption{ Conceptual scheme of the proposed thought experiment for gravitationally induced entanglement (GIE) between two massive particles. Each particle begins in a superposition of two coherent states (a). As the components oscillate and overlap, interference fringes initially appear in the detection probability with full visibility (b–c). Their mutual gravitational interaction generates entanglement over time, leading to reduced fringe visibility at later crossings (d–e).}
    \label{fig:setup}
\end{figure}

We consider two massive particles confined within a three-dimensional harmonic trap, each prepared in a superposition of two coherent states with equal maxiumum displacement proportional to a coherent state parameter \(\alpha \in\mathbb{R} \) (Fig. \ref{fig:setup}). The displacements are chosen along orthogonal axes. The first particle, oscillating along the \( x \)-axis, is initially prepared in a superposition of two coherent states. The first component, labeled \( 1a \), starts at its maximum displacement, while the second component, \( 1b \), is shifted ahead by an eighth of a period, or \( \pi/4 \) oscillations. The second particle oscillates along the \( y \)-axis, with its first component, \( 2a \), beginning at its minimum displacement, where it has its maximum momentum, and is therefore shifted by a quarter period (\(\pi/2\)) relative to \( 1a \). The second component, \( 2b \), is further displaced by another eighth of a period (\(\pi/4\)), making it three-eighths of a period (\(3\pi/4\)) ahead of \( 1a \). Since the particles oscillate along perpendicular axes, these phase relationships ensure that, under the harmonic potential alone, the distance between \( 1a \) and \( 2a \) remains constant, as does the distance between \( 1b \) and \( 2b \).  This initial configuration is shown in Figure \ref{fig:setup}(a) with particle one in red and particle two in blue.

As the system evolves under the harmonic potential, each coherent state component remains approximately Gaussian while following the classical trajectory of a trapped particle. The expectation values of position and momentum for each component oscillate according to the harmonic motion, with the two components of either particle periodically overlapping in position space. These overlaps occur twice per period; for the first particle, whose components oscillate along the \( x \)-axis, they occur at times corresponding to phase angles \( 7\pi/8\), and \( -\pi/8 \). When the two components overlap, interference fringes arise in the probability distribution for detecting the particle's position. Figures \ref{fig:setup}(b) and \ref{fig:setup}(c) show how this interference pattern would appear in the region the components of either particle overlap before entanglement has developed between them.

Due to the gravitational interaction, the evolution of the system leads to entanglement between the two particles. As entanglement grows, the reduced state of either particle deviates from a pure superposition and instead approaches a classical mixture of the two coherent state components. This results in a gradual reduction in the visibility of the interference fringes as shown in Figures \ref{fig:setup}(d) and \ref{fig:setup}(e). To measure this effect, a detector can be placed in one of the regions where the components of a given particle overlap. By performing repeated measurements at different times—corresponding to different numbers of oscillations—the change in fringe visibility can be tracked, providing a direct observable signature of gravitationally induced entanglement.

This setup establishes a well-defined scenario for studying gravitationally induced entanglement in a harmonic trap. The harmonic trap functions as a natural interferometer, allowing entanglement to be probed through measurable changes in fringe visibility as the system evolves. This allows us to model such an interferometric experiment to detect GIE without the use of beam splitters or other various mirrors which would appear as singular potentials in our field theoretic model. Without such impediments we are able to present an operational model for the preparation, evolution and measurement of the masses in this experiment entirely in the language of quantum field theory. In the next section, we take the first step in constructing a field-theoretic model of GIE by developing a relativistic description of the harmonic trap and its coherent states.

\section{\label{sec:scalar field}Scalar Field Model of the Harmonic Trap}

Having established a conceptual framework for gravitationally induced entanglement, we now turn to constructing a field-theoretic model that captures this phenomenon. The previous section described how a pair of massive particles evolve in superpositions of coherent states within a harmonic trap. A complete field-theoretic model of gravitationally induced entanglement must account for both the quantum states of the trapped particles and their gravitational interaction. As a first step, we develop a relativistic framework for the harmonic trap and define coherent states within this setting in an operational manner. We then analyze these coherent states in the weakly relativistic regime to determine the lowest-order corrections to their nonrelativistic counterparts.

To describe this system within quantum field theory, we consider a real scalar field subject to a harmonic trapping potential. The dynamics are governed by the action  
\begin{equation}
    S = \int dt \, d^3x \left[ -\frac{1}{2} \partial_{\mu} \phi \partial^{\mu} \phi - \frac{1}{2} \left(1 + \frac{\omega^2 \bm{x}^2}{c^2} \right) \lambda^{-2} \phi^2 \right].
\end{equation}
We adopt the metric signature \( (-,+,+,+) \) and define the Compton wavelength as \( \lambda = \hbar / mc \). Field theories of this type have been explored in recent studies as models for particle detectors as well as other contexts \cite{oniga_quantum_2017,perche_particle_2024, torres_particle_2024,ragula_localizing_2025}.

To simplify the equations, we introduce the dimensionless variables \(x^{\mu} = \lambda x'^{\mu} = \lambda(t', \bm{x}'), \quad \omega = \frac{c}{\lambda} \omega', \quad \phi = \frac{\sqrt{\hbar c}}{\lambda} \phi'\).
Expressing the action in terms of these variables, we obtain  
\begin{equation}
    S' = \int dt' d^3x' \left[ -\frac{1}{2} \partial'_{\mu} \phi' \partial'^{\mu} \phi' - \frac{1}{2} \left(1 + \omega'^2 \bm{x}'^2 \right) \phi'^2 \right].
\end{equation}
From this point onward, we use these rescaled quantities but omit the primes for simplicity, unless needed for clarity.

The equations of motion for the field are given by  
\begin{equation}
    \ddot{\phi} = \nabla^2\phi - \left(1 + \omega^2 
    \bm{x}^2 \right) \phi.
\end{equation}
To separate the time dependence, we assume a solution of the form  
\(\phi_{\bm{n}} = \Psi_{\bm{n}}(\bm{x}) e^{-i\omega_{\bm{n}} t} + \text{h.c.}\),
where \( \bm{n} = (n_1, n_2, n_3) \). Substituting this into the equation of motion transforms it into an eigenvalue equation of the same form as the Schrödinger equation,  
\begin{equation}
    -\frac{1}{2} \nabla^2 \Psi_{\bm{n}} + \frac{1}{2} \omega^2 \bm{x}^2 \Psi_{\bm{n}} = E_{\bm{n}} \Psi_{\bm{n}},
\end{equation}
with eigenenergies \(E_{\bm{n}} = \frac{1}{2} (\omega_{\bm{n}}^2 - 1)\).

The solutions to this equation correspond to the wavefunctions of a three-dimensional quantum harmonic oscillator, \(\Psi_{\bm{n}}(\bm{x}) = \psi_{n_1}(x_1) \psi_{n_2}(x_2) \psi_{n_3}(x_3)\),
where  
\begin{equation}
    \psi_{n_i}(x_i) = \frac{1}{\sqrt{2^{n_i} n_i!}} \left(\frac{\omega}{\pi}\right)^{1/4} e^{-\frac{\omega x_i^2}{2}} H_{n_i}(\sqrt{\omega} x_i),
\end{equation}
and \( H_{n_i} \) are the Hermite polynomials. The corresponding energy eigenvalues are \(E_{\bm{n}} = \omega \left(|\bm{n}|_1 + \frac{3}{2} \right)\), where $|\bm{n}|_1=n_1+n_2+n_3$ is the one-norm of the multi-index $\bm{n}$
which gives the eigenfrequencies \(\omega_{\bm{n}} = \sqrt{1 + 2\omega (|\bm{n}|_1 +\frac{3}{2})}\).

Using these mode functions, the field operator can be expanded as  
\begin{equation}
    \phi(\bm{x},t) = \sum_{\bm{n}}^{\infty} \frac{1}{\sqrt{2\omega_{\bm{n}}}} \left( a_{\bm{n}} e^{-i\omega_{\bm{n}} t} + a_{\bm{n}}^{\dag} e^{i\omega_{\bm{n}} t} \right) \Psi_{\bm{n}}(\bm{x}).
\end{equation}
The creation and annihilation operators satisfy the standard bosonic commutation relations \([a_{\bm{n}}, a_{\bm{m}}^{\dag}] = \delta_{{\bm{n}}{\bm{m}}}\),
with all other commutators vanishing.

The total momentum operator for the field is defined as \(P^i = -\int d^3x \, \pi \, \partial_i \phi\).
Substituting the mode expansion, we obtain  
\begin{equation}
\begin{aligned}
    P^i &= \frac{i}{2} \sqrt{\frac{\omega}{2}} \sum_{{\bm{n}}=0}^{\infty} \sqrt{n_i + 1} \Bigg(  
    \left( \sqrt{\frac{\omega_{n_i}}{\omega_{{n_i}+1}}} - \sqrt{\frac{\omega_{{n_i}+1}}{\omega_{n_i}}} \right)  
    \left(a_{{n_i}+1} a_{n_i} - a^{\dag}_{{n_i}+1} a^{\dag}_{n_i} \right) \\
    &\quad + \left( \sqrt{\frac{\omega_{n_i}}{\omega_{{n_i}+1}}} + \sqrt{\frac{\omega_{{n_i}+1}}{\omega_{n_i}}} \right)  
    \left(a_{{n_i}+1} a^{\dag}_{n_i} - a^{\dag}_{{n_i}+1} a_{n_i} \right) \Bigg).
\label{eqn:momentum_op}
\end{aligned}
\end{equation}
For clarity in expressing the total momentum operator, we refer only to the relevant component \( n_i \) when it is incremented.

\subsection{\label{subsec:coherent states}Coherent States}

With the scalar field framework for our harmonic trap in place, we now construct $N$-particle coherent states, which serve as the building blocks for the masses in our study of gravitationally induced entanglement. Coherent states in a harmonic potential can be defined in multiple ways, such as through their minimum uncertainty property or their classical-like evolution in phase space. Here, we adopt an operational approach, defining them as displaced ground states, as this corresponds most directly to how they would be prepared in a laboratory setting.

In this field-theoretic context, we construct these states by first creating \( N \) particles in the ground state of the trap and then applying the spatial translation operator \(U(\alpha^{\mu}) = e^{i\mathbf{P} \cdot \boldsymbol{\alpha}}\) with $\bm{P}$ the vector of operators defined in Eq. (\ref{eqn:momentum_op}),
where, without loss of generality, we choose \( \boldsymbol{\alpha} = (\alpha,0,0) \) to simplify expressions by restricting the displacement to a single spatial direction. The resulting \( N \)-particle coherent state is given by  
\begin{equation}
    \ket{\alpha^N} = e^{i\mathbf{P} \cdot \boldsymbol{\alpha}} a_0^{\dag N} \ket{0}.
    \label{eqn:cs}
\end{equation}
To further analyze these states, we define a set of mode operators that will allow us to express the system in a more convenient form. Specifically, we define  
\begin{equation}
    b_{\mathbf{n}} = \int d^3x\, \Psi_{\mathbf{n}}(\bm{x})\left(\phi(\bm{x}) - i\pi(\bm{x})\right).
\end{equation}
This transformation corresponds to a single-mode Bogoliubov transformation of the creation and annihilation operators, preserving the canonical commutation relations. Specifically, \( b_{\mathbf{n}} \) can be written as \(b_{\mathbf{n}} = \mu_{\mathbf{n}} a_{\mathbf{n}} + \nu_{\mathbf{n}} a_{\mathbf{n}}^\dag\), with \(\mu_{\bm{n}}=\frac{1}{2}\left(\sqrt{\omega_{\bm{n}}}+\frac{1}{\sqrt{\omega_{\bm{n}}}}\right)\) and \(\nu_{\bm{n}}=\frac{1}{2}\left(\sqrt{\omega_{\bm{n}}}-\frac{1}{\sqrt{\omega_{\bm{n}}}}\right)\)
where the coefficients satisfy the relation \(|\mu_{\mathbf{n}}|^2 - |\nu_{\mathbf{n}}|^2 = 1\).
Equivalently, \( b_{\mathbf{n}} \) can be expressed as the action of a squeezing operator \( S \) on \( a_{\mathbf{n}} \), \(b_{\mathbf{n}} = S a_{\mathbf{n}} S^\dag\),
where \( S \) is given by  
\begin{equation}
    S = \exp\left(i \sum_{\mathbf{n}=0}^{\infty} \left(\zeta_{\mathbf{n}} a_{\mathbf{n}}^2 - \zeta_{\mathbf{n}}^* a_{\mathbf{n}}^{\dag 2}\right)\right),
\end{equation}
with \( \zeta_{\mathbf{n}} = -\frac{1}{2} \log \omega_{\mathbf{n}} \).  

We introduce another set of operators defined as  
\begin{align}
    b_{\mathbf{p}} &= \sum_{\mathbf{n}=0}^{\infty} \langle \mathbf{n} | \mathbf{p} \rangle b_{\mathbf{n}} = \sum_{\mathbf{n}=0}^{\infty} (-i)^{|\mathbf{n}|_1} \Psi_{\mathbf{n}}(\mathbf{p}) \, b_{\mathbf{n}}, \\
    b_{\mathbf{n}} &= \int d^3p \langle \mathbf{p} | \mathbf{n} \rangle b_{\mathbf{p}} = \int d^3p \, i^{|\mathbf{n}|_1} \Psi_{\mathbf{n}}(\mathbf{p})\, b_{\mathbf{p}}.
\end{align}
Using these, the momentum operator for the field can be written as  
\begin{align}
    P^i &= i\sqrt{\frac{\omega}{2}} \sum_{n=0}^{\infty} \sqrt{n + 1} \left(b_n b^\dag_{n+ i} - b_n^\dag b_{n+ i}\right) \\
    &= \int d^3p\, p^i\, b_{\mathbf{p}}^\dag b_{\mathbf{p}}.
\end{align}
The state in Eq. (\ref{eqn:cs}) can be rewritten as
\begin{equation}
    U(\alpha^{\mu}) a_0^{\dag N} \ket{0} = U S^\dag b_0^{\dag N} S S^\dag \ket{0_b},
\end{equation}
since the squeezing transformation implies that the vacuum of the \( b_{\mathbf{n}} \) modes is related to the original vacuum via $\ket{0_b} = S \ket{0}$. Next we can then transform to the modes $b_{\bm{p}}$ and apply the translation operator by defining a new squeezing operator via $S'=USU^\dag$. This yields
\begin{align}
    U(\alpha^{\mu}) a_0^{\dag N} \ket{0} &= U S^\dag \left(\int d^3p \, \psi_0(\mathbf{p}) \, b_{\mathbf{p}}^\dag \right)^N S S^\dag \ket{0_b} \\
    &= S'^\dag \left(\int d^3p \, \psi_0(\mathbf{p}) \, b_{\mathbf{p}}^\dag e^{-i p_1 \alpha} \right)^N S' S'^\dag \ket{0_b} \\
    &= S'^\dag \left(\sum_{n_1=0}^{\infty} \frac{\alpha^{n_1}}{\sqrt{n_1!}} b_{n_1}^\dag \right)^N S' S'^\dag \ket{0_b}, \\
\end{align}
where the components in the last line can be found by application of  the convolution theorem and we have used the translation invariance of the state $\ket{0_b}$. Now defining new operators via the relation $a_{\bm{n}}'=S'b_{\bm{n}}S'^\dag$, we can write our coherent states as
\begin{equation}
   U(\alpha^{\mu}) a_0^{\dag N} \ket{0} = \left(\sum_{n_1=0}^{\infty} \frac{\alpha^{n_1}}{\sqrt{n_1!}} a_{n_1}'^\dag \right)^N\ket{0'}.
   \label{eqn:RCS}
\end{equation}
These operators relate to the original \( a_{\mathbf{n}} \) via  
\begin{equation}
    a'_{\mathbf{n}} = \frac{1}{2} \left(\sqrt{\frac{\omega_{\mathbf{n}}}{\omega_{n_1=0}}} + \sqrt{\frac{\omega_{n_1=0}}{\omega_{\mathbf{n}}}} \right) a_{\mathbf{n}}
    + \frac{1}{2} \left(\sqrt{\frac{\omega_{\mathbf{n}}}{\omega_{n_1=0}}} - \sqrt{\frac{\omega_{n_1=0}}{\omega_{\mathbf{n}}}} \right) a_{\mathbf{n}}^\dag.
\end{equation}

The field operator can be written in terms of the operators $a_{\bm{n}}'$ as
\begin{equation}
    \begin{split}
    \phi(\bm{x},t) =& \sum_{\bm{n}=0}^\infty \frac{\Psi_{\bm{n}}(\bm{x})}{2\sqrt{2\omega_{\bm{n}}}}\\ & \left( \left( \left(\sqrt{\frac{\omega_{\mathbf{n}}}{\omega_{n_1=0}}} + \sqrt{\frac{\omega_{n_1=0}}{\omega_{\mathbf{n}}}} \right) e^{-i\omega_{\bm{n}}t} - \left(\sqrt{\frac{\omega_{\mathbf{n}}}{\omega_{n_1=0}}} - \sqrt{\frac{\omega_{n_1=0}}{\omega_{\mathbf{n}}}}\right) e^{i\omega_{\bm{n}}t} \right) a_{\bm{n}}' \right. \\
    & \left. + \left( \left(\sqrt{\frac{\omega_{\mathbf{n}}}{\omega_{n_1=0}}} + \sqrt{\frac{\omega_{n_1=0}}{\omega_{\mathbf{n}}}}\right) e^{i\omega_{n_1}t} - \left(\sqrt{\frac{\omega_{\mathbf{n}}}{\omega_{n_1=0}}} + \sqrt{\frac{\omega_{n_1=0}}{\omega_{\mathbf{n}}}} \right) e^{-i\omega_{\bm{n}}t} \right) a_{\bm{n}}^{\dagger'} \right).
    \end{split}
\end{equation}
The expectation value of $\phi^2(x)$ for the state (\ref{eqn:RCS}) is then given by
\begin{equation}
    \begin{split}
        \langle \phi^2 \rangle = & \sum_{\bm{n}=0}^\infty \frac{\Psi_{\bm{n}}(\bm{x})^2}{4} \left|\left(\sqrt{\frac{\omega_{\mathbf{n}}}{\omega_{n_1=0}}} + \sqrt{\frac{\omega_{n_1=0}}{\omega_{\mathbf{n}}}} \right) e^{-i\omega_{\bm{n}}t} - \left(\sqrt{\frac{\omega_{\mathbf{n}}}{\omega_{n_1=0}}} - \sqrt{\frac{\omega_{n_1=0}}{\omega_{\mathbf{n}}}}\right) e^{i\omega_{\bm{n}}t} \right|^2 \\
        & + N \left| \sum_{n_1=0}^\infty \frac{\alpha^{n_1}}{2\sqrt{\omega_{n_1} n_1!}} \psi_{n_1}(x_1) \left( \left(\sqrt{\frac{\omega_{n_1}}{\omega_{n_1=0}}} + \sqrt{\frac{\omega_{n_1=0}}{\omega_{n_1}}} \right) e^{-i\omega_{n_1}t} \right.\right.\\
        &\left.\left.- \left(\sqrt{\frac{\omega_{n_1}}{\omega_{n_1=0}}} - \sqrt{\frac{\omega_{n_1=0}}{\omega_{n_1}}}\right) e^{i\omega_{n_1}t}\right) \right|^2\times\left|\psi_{n_2=0}(x_2)\psi_{n_3=0}(x_3)\right|^2,
    \end{split}
    \label{eqn:expval}
\end{equation}
which is exact for all values of $\alpha, \omega, t$. The first component is a divergent contribution from the time dependent vacuum of the transformed modes, $\ket{0'}$, which would be regularized by the smearing introduced due to the finite size of a particle detector. The second component therefore corresponds to the probability distribution of detecting one of the particles in the multi-particle coherent state at the position $x$. Thus we define the effective one dimensional ``wavefunction" for our relativistic coherent states as

\begin{equation}
\begin{split}
    \psi_{\alpha}(x_1,t)= &\sum_{n_1=0}^\infty \frac{\alpha^{n_1}}{2\sqrt{\omega_{n_1} n_1!}} \psi_{n_1}(x_1) \\
    &\left( \left(\sqrt{\frac{\omega_{n_1}}{\omega_{n_1=0}}} + \sqrt{\frac{\omega_{n_1=0}}{\omega_{n_1}}} \right) e^{-i\omega_{n_1}t} - \left(\sqrt{\frac{\omega_{n_1}}{\omega_{n_1=0}}} - \sqrt{\frac{\omega_{n_1=0}}{\omega_{n_1}}}\right) e^{i\omega_{n_1}t}\right) 
\end{split}  
\label{eqn:RCSwf}
\end{equation}

While our primary interest in this model is as a thought experiment to deepen our fundamental understanding of gravitationally induced entanglement from a first-principles field-theoretic perspective, any potential experimental realization would be limited to probing weakly relativistic corrections to the nonrelativistic limit due to practical constraints. This makes it essential to characterize how these states differ from their familiar nonrelativistic counterparts.

Moreover, we found that in the highly relativistic regime, these states deviate significantly from the usual notion of coherent states, making direct comparisons difficult. To gain insight into their behavior, it is useful to first examine the weakly relativistic limit, where they remain closer to the standard definition. Furthermore, these weakly relativistic states play a central role in our later analysis. When incorporating gravitational interactions and modeling their semiclassical evolution, the approximations we employ are valid in this regime, allowing us to compute analytically tractable results.

To precisely define the weakly relativistic regime, we now identify the relevant physical scales and establish conditions that ensure both the distinguishability of quantum states and the validity of our approximations. The natural confinement length of the trap is given by $x_0 = \sqrt{\frac{\hbar}{m\omega}}$. Here we have used the dimensionful ``unprimed" notation for $\omega$ instead of the rescaled dimensionless values $\omega'$. Since we assume the variations in the probability of detecting the particles are dominated by this length scale, their Compton wavelength must satisfy $\lambda \ll x_0$. Expressed in terms of energy scales, this condition corresponds to $\sqrt{\frac{\hbar \omega}{m c^2}} \ll 1$,
which ensures that the zero-point energy of the potential remains much smaller than the rest energy of the particles. In our dimensionless formulation, this translates to $\sqrt{\omega'} \ll 1$.

Next, for the superposed components of the gravitational harmonium state to be clearly distinguishable, the displacement of the particles, $x_d = \lambda \sqrt{\frac{2}{\omega'}} |\alpha|$, must be large compared to the zero-point uncertainty, implying $x_d \gg x_0$. This condition ensures that the average energy of the displaced state is much larger than the zero-point energy, i.e., $m \omega^2 x_d^2 \gg \hbar \omega$, or equivalently, this requires $|\alpha| \gg 1$.
At the same time, we ensure that particle number non-conserving terms remain negligible by demanding that the total energy remains small compared to the rest energy, $m \omega^2 x_d^2 \ll m c^2$, which translates to $\sqrt{\omega'} |\alpha| \ll 1$.

Finally, to ensure that interference fringes are experimentally accessible, their wavelength must be much larger than the Compton wavelength. Since the fringe spacing is determined by the de Broglie wavelength $\lambda_{dB} = \frac{c \lambda }{\omega x_d}$, the requirement \( \lambda_{dB} \gg \lambda \) also leads to the same constraint,  $m \omega^2 x_d^2 \ll m c^2, \quad \text{or equivalently,} \quad \sqrt{\omega'} |\alpha| \ll 1$. These conditions define the parameter regime  
$x_0 \ll x_d \ll \frac{c}{\omega}$, or equivalently, $1 \ll |\alpha| \ll \frac{1}{\sqrt{\omega'}}$.

To better understand these coherent states in the weakly relativistic regime, we numerically evaluated the probability density of the coherent states by computing the non-vacuum contribution to the expectation value \( \langle \phi^2 \rangle \) [Eq. (\ref{eqn:expval})] in one dimension given by the wavefunction, \(\psi_{\alpha}\) [Eq. (\ref{eqn:RCSwf})]. Since the analytical expression for this quantity involves an infinite sum, we truncated the sum at a finite number of terms for numerical evaluation. Convergence tests showed that for displacement amplitudes up to \( |\alpha| \approx 8 \), the results stabilized as more terms were included in the sum. The dominant contributions to the sum arise from terms near \( n \approx \alpha^2 \), so due to the factorial growth of the terms in the denominator numerical inaccuracies became significant for larger \( \alpha \), limiting the range of reliable computations.

Focusing on the parameter regime where numerical accuracy was maintained, we examined states with relatively large \( \alpha \) and small \( \omega \). In this region, the probability density remained approximately Gaussian over multiple oscillations. However, when plotting the state at integer periods of oscillation, we observed a gradual drift in its position, suggesting a perturbation to the frequency. This indicated a deviation from the expected harmonic motion.

To quantify this effect, we made the ansatz that the leading-order correction to the frequency is proportional to \( -\omega^2 \alpha^2 \), so that the corrected frequency is \(\tilde{\omega} = \omega - \omega^2 \alpha^2\). Replotting the state at intervals corresponding to the corrected period \( 2\pi / \tilde{\omega} \), we found that the drift vanished, confirming that this correction accurately accounts for the observed frequency shift. This result held across a range of different values of \( \omega \) and \( \alpha \), suggesting that it represents a universal lowest-order relativistic correction to the frequency.

\begin{figure}
	\centering
	\begin{subfigure}{.45\textwidth}
		\includegraphics[width=\textwidth]{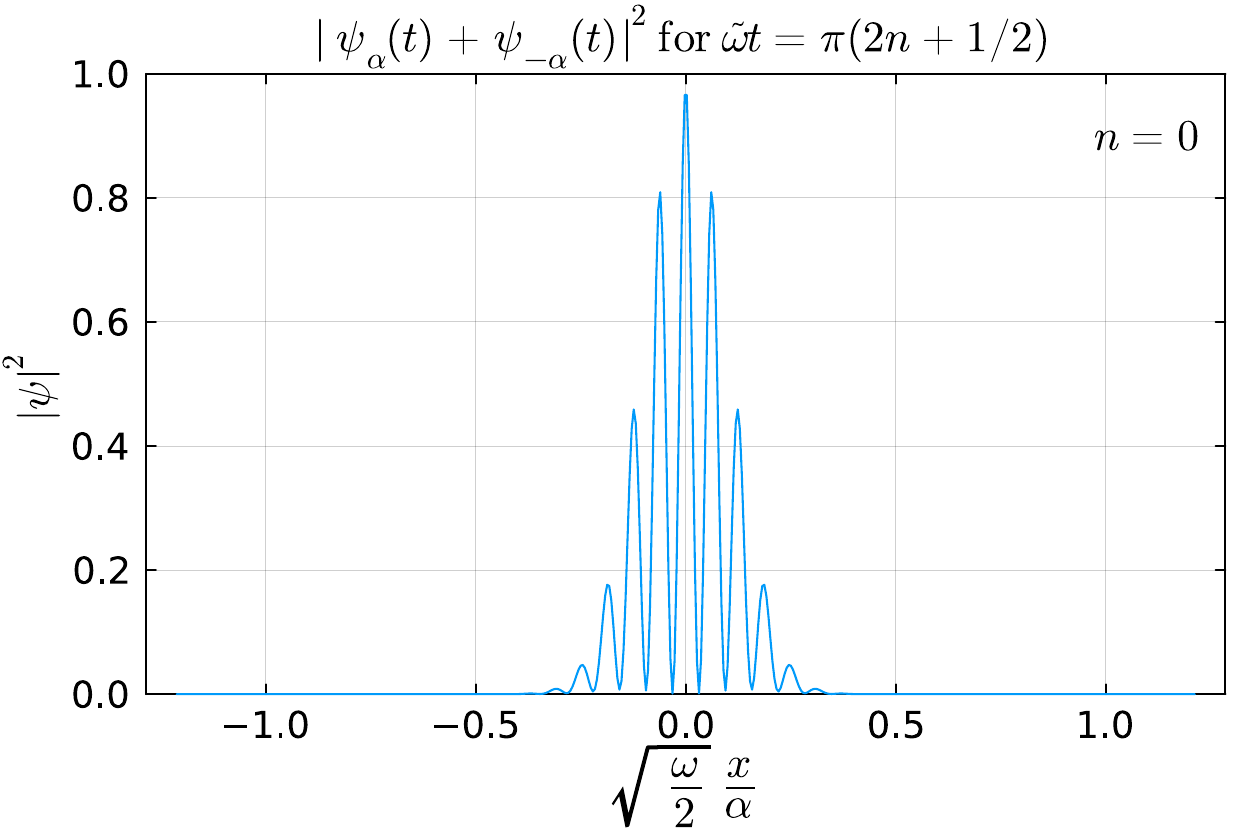}
        \caption{}
	\end{subfigure}
	\begin{subfigure}{.45\textwidth}
		\includegraphics[width=\textwidth]{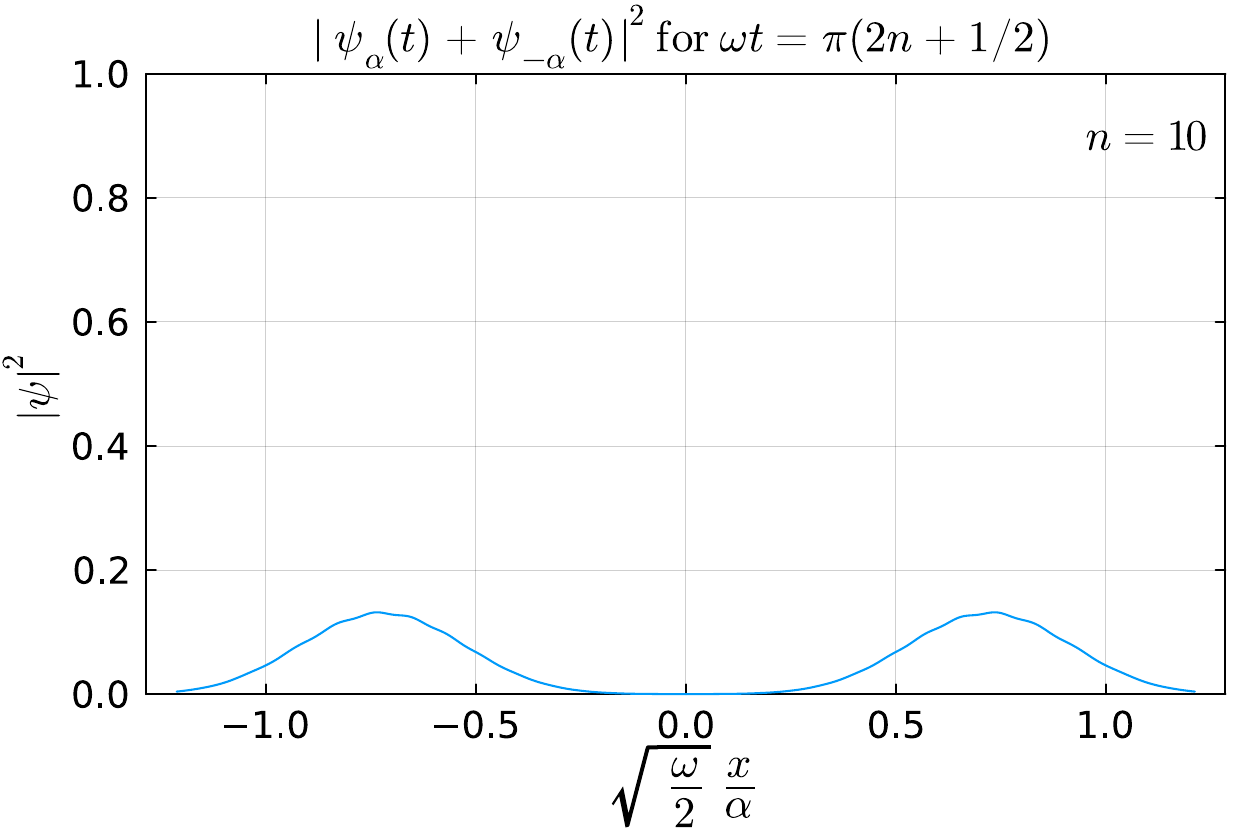}
        \caption{}
	\end{subfigure}
	\begin{subfigure}{.45\textwidth}
		\includegraphics[width=\textwidth]{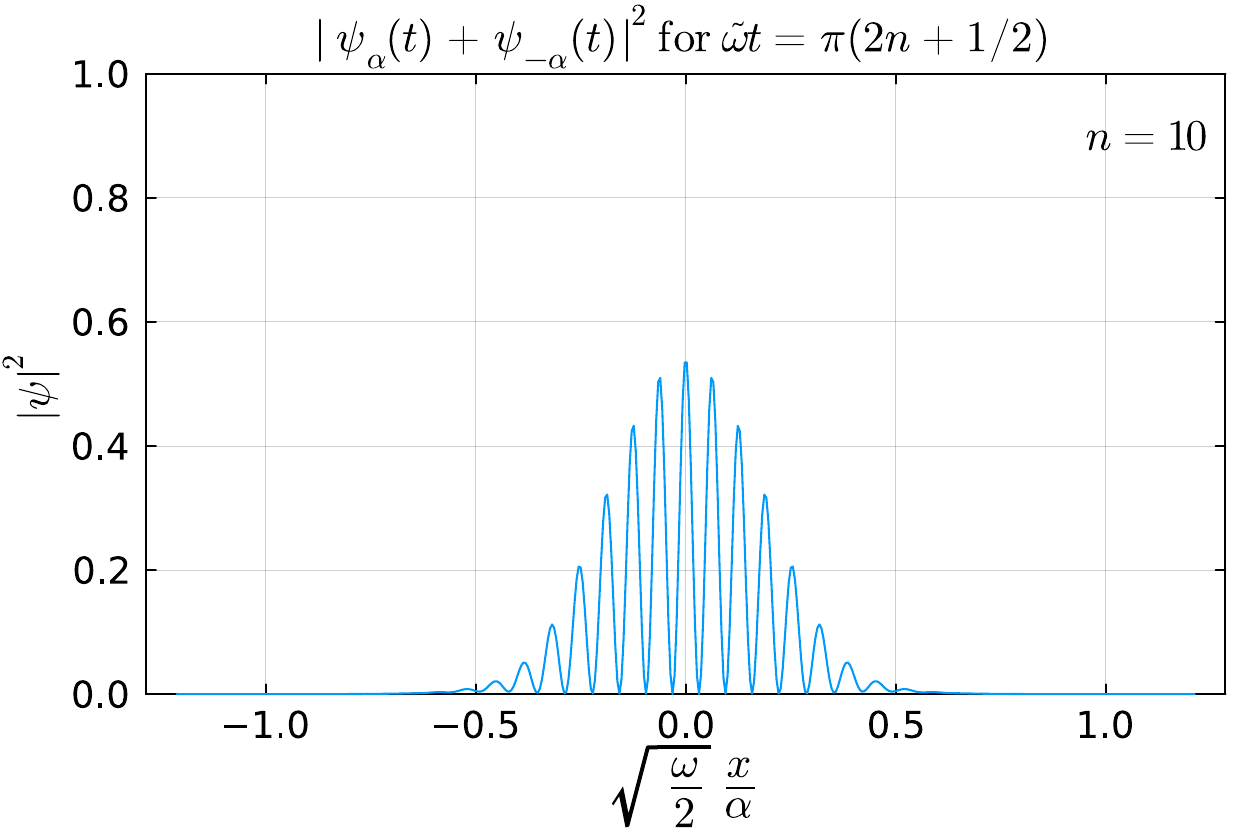}
        \caption{}
	\end{subfigure}
	\caption{Numerical simulation of a superposition of relativistic coherent states with opposite displacements for $\omega = 5 \times 10^{-4}$ and $\alpha = 5$. At the first crossing ($n = 0$), interference fringes appear, closely matching the non-relativistic case (a). By the tenth crossing ($n = 10$), dephasing occurs due to relativistic corrections when evolving with the unperturbed frequency (b). Evolving with the corrected frequency, $\tilde{\omega} = \omega - \omega^2 \alpha^2$, restores overlap. The wavefunctions exhibit broadening from variance modulation, but fringe visibility remains unaffected (c).}
    \label{fig:numerics}
\end{figure}

Further numerical investigations revealed an additional relativistic effect: the variance of the Gaussian probability density was not constant over time. By analyzing the wave packet at various phases of its oscillation across many cycles, we found that the variance was larger near the turning points (where the displacement was maximal) and smaller near the center of the trap. This indicated a periodic modulation of the frequency, suggesting another leading-order relativistic correction. The rate at which these oscillations grew was observed to scale with \( \omega^2 \alpha^2 \), consistent with the first-order frequency shift.

These results confirm that in the weakly relativistic regime, the coherent states remain close to their nonrelativistic counterparts while exhibiting systematic deviations. Importantly, by examining a superposition of coherent states with opposite amplitudes in this regime, we found that the dominant corrections do not affect fringe visibility when the two components overlap; it remains at 100\%, as in the nonrelativistic case. This suggests that relativistic corrections primarily influence the phase evolution of the state rather than the primary observable we intend to use for detecting entanglement. Figure \ref{fig:numerics} demonstrates these numerical results for the case of a superposition of coherent states with opposite initial displacements. Figure \ref{fig:numerics}(a) shows that at the first intersection the probability of particle detection exhibits interference fringes much like NRQM coherent states. Figure \ref{fig:numerics}(b) shows that the states no longer intersect after a time given by several of the expected periods $2\pi/\omega$ due to the shift in the frequency. Figure \ref{fig:numerics}(c) shows that by accounting for the perturbation to the frequency, we can see the snapshots when the components intersect, as well as the broadening of the wavefunctions due to the modulations of the variance, while also confirming that these effects do not diminish the fringe visibility, even though the fringe amplitudes are dimished and the breadth of the fringe pattern is widened.

In this section, we developed a relativistic field-theoretic framework for modeling the quantum states of massive particles in a harmonic trap, establishing the necessary foundation for studying gravitationally induced entanglement in quantum field theory. By defining coherent states in this setting and analyzing their weakly relativistic behavior, we identified leading-order relativistic corrections, including frequency shifts and oscillating variance modulations, which differentiate these states from their nonrelativistic counterparts. Crucially, our numerical investigations confirmed that these corrections do not affect the key observable relevant to our study—fringe visibility—allowing us to proceed with a well-controlled approximation scheme. With this description of the initial states in place, we now turn to the gravitational interaction itself. In the next section, we develop a formalism for linearized quantum gravity in the static limit, providing a field-theoretic description of the gravitational potential that will mediate entanglement between these coherent states.

\section{\label{sec:LQG}The Static Limit of Linearized Quantum Gravity}
Having established a scalar field framework for the harmonic trap and constructed the coherent states that describe the massive particles in our model, we now turn to developing the formalism for their gravitational interaction. In order to incorporate gravity into this field-theoretic description, we employ linearized quantum gravity as an effective theory. To simplify the model further, we focus on the static limit, where we neglect the components of the stress-energy tensor other than  the energy density component. While the static limit is well understood in classical linearized general relativity, its application in linearized quantum gravity—particularly for field-theoretic matter sources—has not been widely explored. Despite the violation of full Lorentz invariance in this approximation, our approach reveals a well-defined causal structure within a chosen reference frame. Additionally, while we neglect higher-order corrections that would restore full relativistic covariance, these corrections can be systematically incorporated into this formalism, providing a natural avenue for future work. In this section, after introducing the framework of linearized gravity, we develop a form of the time evolution operator in the static limit inspired by approaches to similar Hamiltonians in optomechanical systems \cite{bruschi_time_2019,bose_preparation_1997} which have been applied to other models of GIE \cite{bose_spin_2017} (supplementary material). This formulation will provide a clearer understanding of how the gravitational interaction influences the evolution of the matter field and will serve as the foundation for modeling gravitationally induced entanglement in our system.

To describe the gravitational interaction in this framework, we expand the metric around flat spacetime as \(g_{\mu\nu} = \eta_{\mu\nu} + \kappa h_{\mu\nu}\),
where \( h_{\mu\nu} \) represents the quantum gravitational field, and \( \kappa = \frac{\sqrt{32\pi G}}{c^2} \) sets the strength of the coupling.
 The conjugate momentum associated with \( h_{\mu\nu} \) is \(\Pi^{\mu\nu} = \frac{\partial L}{\partial \dot{h}_{\mu\nu}}\),
satisfying the commutation relation  
\begin{equation}
    [h_{\mu\nu}(\bm{x},t), \Pi_{\rho\sigma}(\bm{y},t)] = \frac{i \hbar}{2}(\eta_{\mu\rho} \eta_{\nu\sigma} + \eta_{\mu\sigma} \eta_{\nu\rho}) \delta^3(\bm{x} - \bm{y}).
\end{equation}
The free Hamiltonian of the gravitational field in the De Donder gauge is  
\begin{equation}
    H_{\text{grav}} = \frac{1}{2} \int d^3x \left[  c^2\Pi^{\mu\nu} \Pi_{\mu\nu} +  (\nabla h^{\mu\nu}) \cdot (\nabla h_{\mu\nu}) \right],
\end{equation}
while its interaction with matter is described by  
\begin{equation}
    H_{\text{int}} = -\frac{\kappa}{2} \int d^3x \, h^{\mu\nu} T_{\mu\nu},
\end{equation}
where $T_{\mu\nu}$ is the stress energy tensor of the matter fields. Expanding the metric perturbation in terms of plane waves,
\begin{equation}
    h_{\mu\nu}(\bm{x},t) = \int \frac{d^3p}{\sqrt{2\omega_{\bm{p}}(2\pi \hbar)^{3}}} \left[ c_{\mu\nu}(\bm{p}) e^{i((\bm{p} \cdot \bm{x}  - \omega_{\bm{p}} t))/ \hbar} + c_{\mu\nu}^\dagger(\bm{p}) e^{-i((\bm{p} \cdot \bm{x}  - \omega_{\bm{p}} t))/ \hbar} \right],
\end{equation}
with the relativistic dispersion relation \(\omega_p = \frac{c |p|}{\hbar}\). The creation and annihilation operators obey  
\begin{equation}
    [c_{\mu\nu}(\bm{p}), c_{\rho\sigma}^\dagger(\bm{p}')] = \frac{(2\pi)^3}{2} \delta^3(\bm{p} - \bm{p}') (\eta_{\mu\rho} \eta_{\nu\sigma} + \eta_{\mu\sigma} \eta_{\nu\rho}).
\end{equation}

To simplify the equations, we rescale \(h_{\mu\nu} = \frac{\sqrt{\hbar c}}{\lambda} h'_{\mu\nu}, \quad \kappa = \frac{\lambda}{ \sqrt{\hbar c} }\kappa'\),
and drop the primes for convenience. The rescaled Hamiltonians take the form  
\begin{equation}
    H_{\text{grav}} = \frac{1}{2} \int d^3x \left[ \Pi^{\mu\nu} \Pi_{\mu\nu} + (\nabla h^{\mu\nu}) \cdot (\nabla h_{\mu\nu}) \right],
\end{equation}
\begin{equation}
    H_{\text{int}} = -\frac{\kappa}{2} \int d^3x \, h^{\mu\nu} T_{\mu\nu}.
\end{equation}
In these units, the metric perturbation expands as  
\begin{equation}
    h_{\mu\nu}(\bm{x},t) = \int \frac{d^3p}{\sqrt{2\omega_{\bm{p}}(2\pi )^{3}}} \left[ c_{\mu\nu}(\bm{p}) e^{i(\bm{p} \cdot \bm{x} - \omega_{\bm{p}} t)} + c_{\mu\nu}^\dagger(\bm{p}) e^{-i(\bm{p} \cdot \bm{x} - \omega_{\bm{p}} t)} \right],
\end{equation}
where \( \omega_p = |p| \), and the creation/annihilation operators satisfy  
\begin{equation}
    [c_{\mu\nu}(\bm{p}), c
    _{\rho\sigma}^\dagger(\bm{p}')] = \frac{(2\pi)^3 }{2}\delta^3(\bm{p} - \bm{p}') (\eta_{\mu\rho} \eta_{\nu\sigma} + \eta_{\mu\sigma} \eta_{\nu\rho}).
\end{equation}

To model the gravitational interaction in our system, we now focus on the static limit of linearized quantum gravity. While this approximation simplifies the problem by isolating the dominant energy density contribution to the gravitational interaction, our primary motivation for adopting this limit is that it enables analytical progress, allowing us to derive an effective Hamiltonian that governs the evolution of the matter field. This effective description provides a tractable framework for approximating the time evolution of the states used in our gravitationally induced entanglement setup. Importantly, although we neglect the dynamical components of the stress-energy tensor, our approach remains compatible with a relativistic source for the energy density, which we will exploit in later sections.

Within this limit, the gravitational field is sourced only by the energy density \( \rho = T^{00} \), and the metric perturbation reduces to a single degree of freedom, \( h_{00} \), which we denote simply as \( h \) and we denote the mode operators simply by \(c_p\). Under these assumptions, the total Hamiltonian governing the coupled matter and gravitational fields takes the form:
\begin{equation}
    H = \int d^3x \rho + \int d^3p\, \omega_{\bm{p}}\, c_{\bm{p}}^\dag c_{\bm{p}} + \frac{\kappa}{2} \int d^3x \rho h.
    \label{eqn:static Hamiltonian}
\end{equation}
We then rewrite Eq. (\ref{eqn:static Hamiltonian}) using the expansion of \( h \), leading to:
\begin{equation}
    H = \int d^3x \rho + \int d^3p\, \omega_{\bm{p}}\, \left((c_{\bm{p}}^\dag + \beta_{\bm{p}}^\dag)(c_{\bm{p}} + \beta_{\bm{p}}) - \beta_{\bm{p}}^\dag \beta_{\bm{p}}\right),
\end{equation}
where \(\beta_{\bm{p}} = \frac{\kappa}{2\sqrt{2\omega_{\bm{p}}^3(2\pi)^3}} \int d^3x \rho e^{-i{\bm{p}}\cdot\bm{x}}\).

To simplify, we introduce new operators:
\begin{equation}
    \mathcal{D}(\{\beta_{\bm{p}}\}) = \exp\left(\int d^3p (\beta_{\bm{p}} c_{\bm{p}}^\dag - \beta_{\bm{p}}^\dag c_{\bm{p}})\right),
\end{equation}
which displace the graviton vacuum. Using this, we rewrite \( H \) as
\(H = \mathcal{D}^\dag(\{\beta_{\bm{p}}\}) \tilde{H} \mathcal{D}(\{\beta_{\bm{p}}\})\),
where the displaced Hamiltonian is:
\begin{equation}
    \tilde{H} = \int d^3x \rho + \int d^3p\, \omega_{\bm{p}} (c_{\bm{p}}^\dag c_{\bm{p}} - \beta_{\bm{p}}^\dag \beta_{\bm{p}}),
\end{equation}
This allows us to write the evolution operator as \(e^{-iHt} = D^\dagger(\{\beta_{\bm{p}}\}) e^{-i\tilde{H}t} D(\{\beta_{\bm{p}}\})\).
To simplify, we use the identity \(e^{-i\tilde{H}t} D(\{\beta_{\bm{p}}\}) e^{i\tilde{H}t} = D(\{\beta_{\bm{p}} e^{i\omega_{\bm{p}} t}\})\). Substituting this into our expression, we obtain  
\begin{equation}
    e^{-iHt} = e^{-i\tilde{H}t} D^\dagger(\{\beta_{\bm{p}} e^{i\omega_{\bm{p}} t}\}) D({\beta_{\bm{p}}}).
\end{equation}
Since the product of displacement operators satisfies  
\begin{equation}
    D^\dagger(\{\beta_{\bm{p}} e^{i\omega_{\bm{p}} t}\}) D(\{\beta_{\bm{p}}\}) = \exp\left(-i \int d^3p \, \beta_{\bm{p}}^\dagger \beta_{\bm{p}} \sin \omega_{\bm{p}} t \right) D(\{\beta_{\bm{p}}(1 - e^{i\omega_{\bm{p}} t})\}),
\end{equation}
we now express the exponent in terms of the matter energy density by recalling the definition of \( \beta_{\bm{p}} \):  
\begin{equation}
    \int d^3p \, \beta_{\bm{p}}^\dagger \beta_{\bm{p}} \sin \omega_{\bm{p}} t = \frac{\kappa^2}{32\pi} \int d^3x d^3y \, \rho(\bm{x}) \rho(\bm{y}) \Theta(t - |\bm{x} - \bm{y}|) \left( \frac{t}{|\bm{x} - \bm{y}|} - 1 \right).
\end{equation}
Additionally, the remaining displacement operator can be rewritten in terms of the freely evolved gravitational field as  
\begin{equation}
    D(\{\beta_{\bm{p}}(1 - e^{i\omega_{\bm{p}} t})\}) = \exp\left(-i\frac{\kappa}{2} \int_0^t dt' \int d^3x \rho(\bm{x}) h_0(\bm{x}, t') \right).
\end{equation}
Substituting this result, the evolution operator takes the final form:  
\begin{equation}
    \begin{split}
        e^{-iHt} =& \exp\left\{ -i \left[ \int d^3x \rho(\bm{x}) t + \int d^3p \, \omega_{\bm{p}} t \, c_{\bm{p}}^\dagger c_{\bm{p}} \right. \right. \\
        &\left. \left. - \frac{\kappa^2}{32\pi} \int d^3x d^3y \, \rho(\bm{x}) \rho(\bm{y}) \Theta(t - |\bm{x} - \bm{y}|) 
        \left( \frac{t}{|\bm{x} - \bm{y}|} - 1 \right) \right] \right\}\\
        &\exp\left(-i\frac{\kappa}{2} \int_0^t dt' \int d^3x \rho(\bm{x}) h_0(\bm{x}, t') \right).
    \end{split}
\end{equation} 
We note that neglecting the components of the stress-energy tensor beyond the energy density not only simplifies the interaction term in the Hamiltonian but also effectively neglects the time evolution of the energy density itself, which would otherwise be generated by the $T^{0i}$ components.

Neglecting the entanglement between the matter and gravitational field corresponds to expanding the displacement term,  
\begin{equation}
\exp\left(-i\frac{\kappa}{2} \int_0^t dt' \int d^3x \rho(\bm{x}) h_0(\bm{x}, t') \right),
\end{equation}  
to zeroth order, which would otherwise contribute to decoherence in measurements that probe only the matter degrees of freedom. After making this approximation, tracing out the gravitational field yields unitary evolution of the matter fields under the operator  
\begin{equation}
\exp\left\{-i\left[\int d^3x \rho(\bm{x}) t - \frac{\kappa^2}{32\pi} \int d^3x d^3y \, \rho(\bm{x}) \rho(\bm{y}) \Theta(t - |\bm{x} - \bm{y}|) \left(\frac{t}{|\bm{x} - \bm{y}|} - 1 \right) \right]\right\}.
\end{equation}  
This corresponds to evolution under an effective Hamiltonian,  
\begin{equation}
    H_{\text{eff}}(t) = \int d^3x \rho(x) - \frac{\kappa^2}{32\pi} \int d^3x d^3y \, \Theta(t - |\bm{x} - \bm{y}|) \frac{\rho(\bm{x}) \rho(\bm{y})}{|\bm{x} - \bm{y}|},
    \label{eqn:H_eff}
\end{equation}  
with the time evolution operator \(e^{-i\int_0^t dt' H_{\text{eff}}(t')}\). 

In this section, we have derived the time evolution operator for the coupled matter-gravity system in the static limit of linearized quantum gravity. By expressing the evolution in terms of a displacement transformation, we identified how the gravitational field shifts under the influence of the matter energy density. Tracing out the gravitational degrees of freedom led to an effective Hamiltonian that governs the dynamics of the matter field, capturing the leading-order effects of gravity in this limit.

This result provides a powerful framework for analyzing gravitationally induced entanglement in a harmonic trap. The effective Hamiltonian dynamics allow us to approximate the evolution of quantum states while systematically incorporating relativistic corrections to the gravitational interaction. In the next section, we apply these results to our field-theoretic model of GIE, examining how the gravitational interaction influences the coherent states constructed earlier and how this effect can be probed experimentally.

\section{\label{sec:QFT GIE}QFT model of GIE}

Our goal is to develop a fully field-theoretic model of gravitationally induced entanglement, providing a framework that incorporates both the dominant relativistic corrections to the trapped masses' dynamics and their gravitational interaction. With the field-theoretic description of the harmonic trap and coherent states established, along with the effective Hamiltonian dynamics derived from linearized gravity, we now apply these methods to the conceptual setup defined in Sec. \ref{sec:set up} to approximate the system’s time evolution. We then compute the probability distribution for detecting one of the masses by evaluating expectation values of field observables and use this to track the decay of fringe visibility in the overlap region of the mass’s two coherent state components.

To implement this approach, we first express the effective Hamiltonian in terms of the scalar field energy density. The energy density takes the form  

\begin{equation}
    \rho = \frac{1}{2} \sum_{\bm{n},\bm{m}=0}^\infty (\omega_{\bm{n}} \omega_{\bm{m}})^{-1/2} \left[ \nabla \psi_{\bm{n}} \cdot \nabla \psi_{\bm{m}} + (1+\omega^2 \bm{x}^2 + \omega_{\bm{n}} \omega_{\bm{m}}) \psi_{\bm{n}} \psi_{\bm{m}} \right] a^\dag_{\bm{n}} a_{\bm{m}}.
\end{equation}  
Substituting this into the effective Hamiltonian [Eq. (\ref{eqn:H_eff})] and applying the rotating wave approximation (RWA) to neglect particle number non-conserving terms, we obtain  
\begin{equation}
\begin{split}
    H_{\text{eff}}(t) = & \sum_{\bm{n}=0}^\infty \omega_{\bm{n}} a_{\bm{n}}^\dag a_{\bm{n}} \\
    & -\frac{ \kappa^2}{32\pi} \sum_{\bm{n},\bm{m},\bm{j},\bm{k=}0}^\infty \int d^3x d^3y \, \psi_{\bm{n}}(\bm{x}) \psi_{\bm{m}}(\bm{x}) \psi_{\bm{j}}(\bm{y}) \psi_{\bm{k}}(y) (\omega_{\bm{n}} \omega_{\bm{m}} \omega_{\bm{j}} \omega_k)^{-1/2} \\
    & \quad \times \left( (\omega_{\bm{n}}+\omega_{\bm{m}})^2 + \nabla_{\bm{x}}^2 \right) \left( (\omega_{\bm{j}}+\omega_{\bm{k}})^2 + \nabla_{\bm{y}}^2 \right) \frac{\Theta(t - |\bm{x} - \bm{y}|)}{|\bm{x} - \bm{y}|} a_{\bm{n}}^\dag a_{\bm{m}} a_{\bm{j}}^\dag a_{\bm{k}}.
\end{split}
\end{equation}

To further analyze the effective Hamiltonian, we express it in a first-quantized form that acts on the Hilbert space of an \( N \)-particle system. This isomorphism is established by identifying the operator whose matrix elements match those of \( H_{\text{eff}}(t) \) in a basis given by the harmonic oscillator eigenstates of an \( N \)-particle system.

For the free part of the Hamiltonian, the term \(\sum_{\bm{n}=0}^\infty \omega_{\bm{n}} a_{\bm{n}}^\dag a_{\bm{n}}\)
describes the sum of harmonic mode energies, where the mode frequencies are given by \(\omega_{\bm{n}} = \sqrt{1 + 2\omega(|\bm{n}|_1 + 3/2)}\).
In an \( N \)-particle system, this has the same matrix elements as the operator \(\sum_{j=1}^{N} \sqrt{1 + P_j^2 + \omega^2 X_j^2}\),
which sums the single-particle relativistic energy contributions for each particle. A similar correspondence can be made for the gravitational interaction term, mapping it to an operator acting on the \( N \)-particle Hilbert space. We express the effective Hamiltonian in first-quantized notation as  

\begin{equation}  
H_{\text{eff}}(t) = \sum_{j=1}^N \sqrt{1 + P_j^2 + \omega^2 X_j^2} + \sum_{\substack{j,k=1 \\ j \neq k}}^N V(P_j, P_k, X_j, X_k, t),
\label{eqn:1st quantized}
\end{equation}  
where \( V(P_j, P_k, X_j, X_k, t) \) represents the gravitational interaction.

The initial state of our matter system is given by a product of two \(N\)-particle Schrödinger cat states. Each cat state consists of a superposition of two  \(N\)-particle coherent states:  
\begin{equation}
    \ket{\psi(0)} = \mathcal{N}\left(\ket{\alpha_{1a}^N} + \ket{\alpha_{1b}^N}) \otimes (\ket{\alpha_{2a}^N} + \ket{\alpha_{2b}^N}\right)
\end{equation}
\begin{equation}
    =\mathcal{N}\left( \ket{\alpha_{1a}^N \alpha_{2a}^N} + \ket{\alpha_{1a}^N \alpha_{2b}^N} + \ket{\alpha_{1b}^N \alpha_{2a}^N} + \ket{\alpha_{1b}^N \alpha_{2b}^N}\right),
\end{equation}
where $\mathcal{N}$ is a normalization factor. The coherent state parameters are given by \(\alpha_{1a} = \alpha, \quad \alpha_{1b} = \alpha e^{i\pi/4}, \quad \alpha_{2a} = \alpha e^{i\pi/2}, \quad \alpha_{2b} = \alpha e^{i3\pi/4}\),
where the components of mass 1 are displaced along the \(x\)-axis, and the components of mass 2 are displaced along the \(y\)-axis, consistent with the initial conditions described in Sec. \ref{sec:set up}.

To approximate the time evolution of this system under the effective Hamiltonian, we adapt a semiclassical approximation scheme from Ref. \cite{heller_time-dependent_1975}. Since each component \(\ket{\alpha_{1j}^N \alpha_{2k}^N}\) is a product of coherent states, its wavefunction is well approximated by a Gaussian in the weakly relativistic regime. In this method, the Hamiltonian is expanded to second order around the expectation value of the position operator of one of these components, yielding equations of motion for the Gaussian parameters. However, because our first-quantized Hamiltonian is not quadratic in the momentum operators, we extend this approach by additionally expanding around the momentum expectation value. The resulting equations of motion allow us to treat the gravitational interaction as a perturbation to the harmonic trapping potential.

A similar approximation was employed in our previous work \cite{yant_gravitationally_2023}, where we studied two particles in a nonrelativistic harmonic trap interacting via Newtonian gravity. There, we found that the dominant contribution to gravitationally induced entanglement arises from the relative phase evolution of the Gaussian components of the initial state. These components were well approximated by coherent states evolving under the harmonic oscillator potential, with an additional phase term due to the gravitational interaction. The semiclassical equations of motion for the Gaussian parameters yield a phase shift given by the classical action of the full, unexpanded Hamiltonian along the trajectory of the expectation values of position and momentum. 

Expanding this action to first order in the gravitational interaction, the leading-order term corresponds to the classical action along the unperturbed trajectory and is already encoded in the evolution of the coherent state. Therefore, the time evolution of each component of the initial state can be approximated by  
\begin{equation}
    e^{-i\delta S_{jk}^{(1)}}\ket{\alpha_{1j}^N \alpha_{2k}^N(t)},
\end{equation}
where \(\ket{\alpha_{1j}^N \alpha_{2k}^N(t)}\) represents a time-evolved product of coherent states, while the phases \(\delta S^{(1)}_{jk}\) correspond to the first-order perturbation to the classical action of \(H_{\text{eff}}\) [Eq. (\ref{eqn:1st quantized})] evaluated along the expectation value trajectories. We extend this approximation scheme to each of the four components by linearity. The approximate time evolution of the state is given by  
\begin{equation}
\begin{split}
    \ket{\psi(t)} = &\mathcal{N}\left(e^{-i\delta S_{aa}^{(1)}}\ket{\alpha_{1a}^N \alpha_{2a}^N(t)} + e^{-i\delta S_{ab}^{(1)}}\ket{\alpha_{1a}^N \alpha_{2b}^N(t)}\right.\\
    &\left.+ e^{-i\delta S_{ba}^{(1)}}\ket{\alpha_{1b}^N \alpha_{2a}^N(t)} + e^{-i\delta S_{bb}^{(1)}}\ket{\alpha_{1b}^N \alpha_{2b}^N(t)}\right).
\end{split}
\label{eqn:final state}
\end{equation}

It is important to note that the expansion of the effective Hamiltonian around expectation values is not valid for the gravitational interaction terms between particles within the same coherent state component as the gravitational potential between them diverges along their expectation values. Since all particles in each component occupy identical states to a good approximation, these self-interaction terms contribute only an overall factor, which does not affect the relative phase evolution of the superposition components. As such, they can be neglected in the following analysis.

To evaluate the classical actions, we treat the gravitational interaction as a perturbation to the harmonic potential. The classical version of the effective Hamiltonian is expressed as \(H_{\text{eff}} = H_0 + \Delta H\),
where the unperturbed Hamiltonian describes the relativistic harmonic trap energies,  
\begin{equation}
    H_0 = \sum_{j=1}^{N} H_{0j} = \sum_{j=1}^{N} \sqrt{1 + P_j^2 + \omega^2 X_j^2},
    \label{eqn:H_0}
\end{equation}
and the perturbation due to the gravitational interaction is given by  
\begin{equation}
    \Delta H = -\frac{\kappa^2}{32\pi} \sum_{\substack{j,k=1 \\ j \neq k}}^N \frac{H_{0j} H_{0k}}{|X_j - X_k|}.
    \label{eqn:DH}
\end{equation}
This expression corresponds to the classical version of the gravitational interaction term in Eq. (\ref{eqn:1st quantized}), which is simplified in the form presented here due to the commutativity of terms at the classical level. Additionally, we have omitted the step function and Laplacian terms present in the full interaction Hamiltonian. This simplification is justified because the timescales over which we evolve the system are much longer than the light-crossing time between the masses. Furthermore, the Laplacian terms involve higher negative powers of the interparticle distances, making them subdominant. Any remaining contributions proportional to delta functions contribute only small constant terms to the perturbation of the action and can thus be neglected. These terms represent the causal nature of the interaction derived in Sec. \ref{sec:LQG} which are negligible on the scales that would be involved in a practical implementation of this set up.

In the remainder of this work, we restrict our analysis to two spatial dimensions, neglecting the third since its dynamics are trivial. Each particle’s position and momentum, \(X_j\) and \(P_j\), are two-dimensional vectors, whose components along the Cartesian axes we denote as \(X_{j1}, X_{j2}\) and \(P_{j1}, P_{j2}\).  

The states considered here describe \(2N\)-particle systems, where the first \(N\) particles compose the first mass and the remaining \(N\) particles form the second mass. To reflect the initial conditions of each component of the state, we set \(X_{j2} = P_{j2} = 0\) for particles in the first mass and \(X_{j1} = P_{j1} = 0\) for those in the second mass. This configuration ensures that the first mass is initially displaced along the \(x\)-axis while the second mass is displaced along the \(y\)-axis, consistent with the setup defined in Sec. \ref{sec:set up}.

To proceed, we apply classical canonical perturbation theory to this form of the effective Hamiltonian. We first transform to the Hamilton-Jacobi coordinates of the unperturbed system, which consist of a set of canonically conjugate constants of motion. The Hamilton-Jacobi coordinates of \( H_0 \) [Eq. (\ref{eqn:H_0})], denoted \( \beta_j \) and \( \gamma_j \), are given by
\begin{align} 
    X_{j1}(t) &= \frac{\sqrt{2\gamma_{j1}}}{\omega} \sin\left(\frac{\omega}{H_{0j}}t+\omega\beta_{j1}\right), \\ 
    P_{j1}(t) &= \sqrt{2\gamma_{j1}} \cos\left(\frac{\omega}{H_{0j}}t+\omega\beta_{j1}\right),
\end{align} 
with analogous expressions for \( X_{j2} \) and \( P_{j2} \).

In these coordinates, the unperturbed Hamiltonian [Eq. (\ref{eqn:H_0})] vanishes, leaving only the perturbative potential [Eq. (\ref{eqn:DH})]. To determine its effect, we solve the equations of motion iteratively to first order. The first-order corrections satisfy
\begin{align} 
    \dot{\beta}_{j}^{(1)} &= \left. \frac{\partial \Delta H}{\partial \gamma_{j}} \right|_0, \\ 
    \dot{\gamma}_{j}^{(1)} &= -\left. \frac{\partial \Delta H}{\partial \beta_{j}} \right|_0,
\end{align}
where the right-hand side is evaluated along the unperturbed zeroth-order trajectories.

To establish a reference for comparison, we first recall the nonrelativistic version of this problem, where two particles in a harmonic trap interact via Newtonian gravity. The Hamiltonian is given by  
\begin{equation}
    H = \sum_{j=1}^{2} \frac{1}{2} P_j^2 + \frac{1}{2} \omega^2 X_j^2 - \frac{\kappa^2}{32\pi |X_1 - X_2|},
\end{equation}
with Hamilton-Jacobi coordinates related by  
\begin{align}
    X_j(t) &= \frac{\sqrt{2\gamma_{xj}}}{\omega} \sin\left(\omega t + \omega \beta_{xj} \right),  \\
    P_{xj}(t) &= \sqrt{2\gamma_{xj}} \cos\left(\omega t + \omega \beta_{xj} \right).  
\end{align}
The first-order correction to the classical action in this case is given by  
\begin{equation}
    \begin{split}
        \frac{\kappa^2}{32\pi} \frac{1}{\sqrt{2\gamma}} 
        \int_0^{s'} \int_0^{s_1} 
        \left( \sin^2(\beta_x + s_2) + \sin^2(\beta_y + s_2) \right)^{-3/2}  
        f(\beta_x, \beta_y, s_1, s_2) \, ds_2 \\
        + \frac{1}{\sqrt{\sin^2(\beta_{j1} + s_1) + \sin^2(\beta_{k2} + s_1)}} ds_1.
    \end{split}
    \label{eqn:newtonian ds1}
\end{equation}
where \( s' = \omega t \), and 
\begin{equation}
\begin{split}
    f(\beta_x, \beta_y, s_1, s_2) = &\left( \sin(2\beta_{j1} + 2s_1) \sin^2(\beta_{j1} + s_2) 
    + \frac{1}{2} \cos(2\beta_{j1} + 2s_1) \sin(2\beta_{j1} + 2s_2) \right. \\
    &+ \sin(2\beta_{k2} + 2s_1) \sin^2(\beta_{k2} + s_2) 
    + \frac{1}{2} \cos(2\beta_{k2} + 2s_1) \sin(2\beta_{k2} + 2s_2) \\
    &+ \left. \frac{1}{2} \cos(2\beta_{k2} + 2s_1) \sin(2\beta_{k2} + 2s_2) \right).
\end{split}
\end{equation}

In the relativistic case, we find that the first-order perturbation to the action consists of two contributions: one that resembles the Newtonian result [Eq. (\ref{eqn:newtonian ds1})], denoted \( \delta S^{(1)\,NR} \), and an additional relativistic contribution, \( \delta S^{(1)\,rel} \). These are given by  

\begin{equation}
\begin{split}
    \delta S^{(1)\,NR} =& \frac{\kappa^2}{32\pi} \frac{(1+2\gamma)^{3/2}}{\sqrt{2\gamma}} \times\\
    &\sum_{j=1,k=N+1}^{N,2N} \int_0^s \int_0^{s_1}f(\beta_x, \beta_y, s_1, s_2) \left( \sin^2(\beta_{j1} + s_2) + \sin^2(\beta_{k2} + s_2) \right)^{-3/2} ds_2 \\
    &\quad + \frac{1}{\sqrt{\sin^2(\beta_{j1} + s_1) + \sin^2(\beta_{k2} + s_1)}} ds_1.
\end{split}
\label{eqn:ds1NR}
\end{equation}
The additional relativistic contribution \( \delta S^{(1)\,rel} \) arises from relativistic corrections due to the coupling of the relativistic energy of the harmonic trap to both the trajectory's frequency and the effective gravitational interaction. This contribution is given by  

\begin{equation}
\begin{split}
    \delta S^{(1)\,rel} =& \frac{\kappa^2}{32\pi} \sqrt{2\gamma} \sqrt{1+2\gamma} \;\times\\
    &\sum_{j=1,k=N+1}^{N,2N} \int_0^s \int_0^{s_1} \frac{1}{2} \left( \left( \sin (2\beta_{j1} + 2 s_2)\right.\right.\\
    &\left.\left.\left( (s_2 - s_1) \sin (2 \beta_{j1} + 2 s_1)- \cos^2(\beta_{j1} + s_1) \right)\right. \right.  \\
    & \left. \left.  + \sin (2 \beta_{k2} + 2 s_2) \left( (s_2 - s_1) \sin (2\beta_{k2} + 2 s_1) - \cos^2(\beta_{k2} + s_1) \right) \right) \right) \\
    &\quad \times \left( \sin^2(\beta_{j1} + s_2) + \sin^2(\beta_{k2} + s_2) \right)^{-3/2}\\
    &+ \left( \frac{\sin (2\beta_{j1} + 2 s_1) - \sin (2\beta_{k2} + 2 s_1)}{\sqrt{\sin^2(\beta_{j1} + s_2) + \sin^2(\beta_{k2} + s_2)}} \right)ds_2ds_1.
\end{split}
\label{eqn:ds1rel}
\end{equation}
These results are expressed in terms of the zeroth-order values of all \( \beta \) and \( \gamma \), which specify the initial conditions. We have rescaled \( \beta \rightarrow \beta/\omega \) and assumed that all particles share the same maximum displacement amplitude. Consequently, the values of \( \gamma_{j1} \) for the first mass and \( \gamma_{j2} \) for the second mass are equal, denoted simply by \( \gamma \). Additionally, we introduce the dimensionless parameter \(s = \frac{\omega t}{\sqrt{1+2\gamma}}\).

Next, we observe that the first-order correction to the action can be expressed in the form  
\begin{equation}
    \delta S^{(1)} = \frac{\kappa^2}{32\pi} \left( \frac{(1+2\gamma)^{3/2}}{\sqrt{2\gamma}} I^{NR} \left( \frac{\omega t}{\sqrt{1+2\gamma}} \right) 
    + \sqrt{2\gamma} \sqrt{1+2\gamma} I^{rel} \left( \frac{\omega t}{\sqrt{1+2\gamma}} \right) \right),
\end{equation}
where \( I^{NR} \) and \( I^{rel} \) represent integrals implicitly defined by Eqs. (\ref{eqn:ds1NR}) and (\ref{eqn:ds1rel}). Additionally, for the coherent states considered here, the parameter \( \gamma \) is related to the coherent state amplitude by \(\gamma = \omega \alpha^2\). As shown in Sec. \ref{sec:scalar field}, the weakly relativistic regime corresponds to \( \omega \alpha^2 \ll 1 \). Expanding the correction to next-to-leading order in this parameter, we obtain  
\begin{equation}
\begin{split}
    \delta S^{(1)} = &\frac{\kappa^2}{32\pi} \left( \frac{1}{\sqrt{2\omega} \alpha} I^{NR}\left(\omega t\right)+ \sqrt{2\omega} \alpha \left( \frac{3}{2} I^{NR}(\omega t) + I^{rel}(\omega t)-\frac{1}{2}\dot{I}^{NR}(\omega t) \right) \right).
\end{split}
\label{eqn:ds1wr}
\end{equation}
This result explicitly separates the leading-order contribution, which corresponds to the nonrelativistic interaction, from the next-to-leading order relativistic corrections. The first term, scaling as \( 1/\alpha \), matches the expected form of the nonrelativistic action correction, while the second term introduces relativistic modifications that depend on both \( I^{NR} \) and \( I^{rel} \).

For the final state \( \ket{\psi(t)} \) with the actions derived above, we propose measuring the fringe visibility in the interference pattern within the region where the two coherent state components of one of the masses overlap. In our previous work, we demonstrated that in the nonrelativistic version of this system, a decrease in fringe visibility correlated with an increase in entanglement between the two masses. Since our model neglects all other potential sources of fringe visibility loss, we take this effect as a signature of gravitationally induced entanglement.  

To quantify the fringe visibility, we require the probability distribution for detecting one of the masses, which can be described using the expectation value of an observable constructed from matter field operators. This observable must probe the position of one of the masses, which, as a composite system of many identical particles, is effectively described by its center of mass. Since all particles have equal mass, the center of mass is simply their average position.  

In Sec. \ref{sec:scalar field}, we noted that the expectation value of the squared field strength, \( \phi^2(x) \), in Eq. (\ref{eqn:expval}), is analogous to the probability density of detecting a single particle. The probability density for detecting the center of mass of a collection of particles at a given location corresponds to detecting all particles in a configuration where their average position matches that location. Based on this, we propose the following observable:  
\begin{equation}
    \phi^2_R = \;:\int_{R=\frac{1}{N}\sum_{j=1}^N x_j} \prod_{j=1}^N dx_j \phi^2(x_j):,
\end{equation}
where $::$ denotes normal ordering in which all creation operators are moved to the left and annihilation operators to the right. This effectively measures the probability density of the center of mass of $N$ particles. Neglecting the vacuum contribution, in the weakly relativistic regime, this observable can be well approximated by  
\begin{equation}
    \phi^2_R \approx \ket{R} \bra{R} + \mathcal{O}(\omega),
\end{equation}  
where \( \ket{R} \bra{R} \) is the projector onto the center-of-mass coordinate in the first-quantized representation.

\begin{figure}
    \centering
    \includegraphics[width=0.5\linewidth]{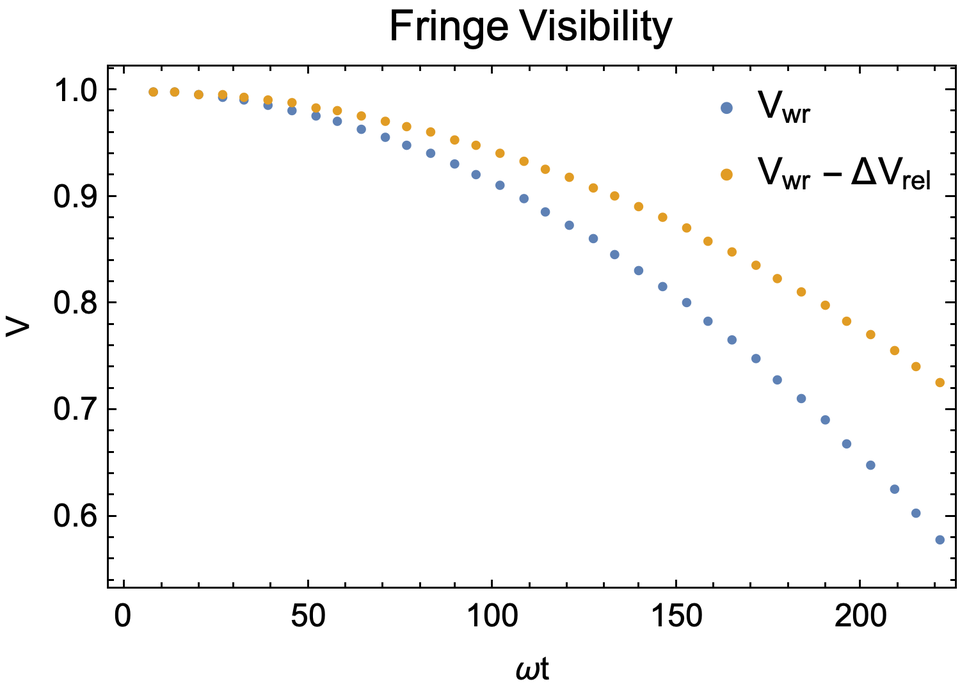}
    \caption{Fringe visibility $V_{wr}$ in the weakly relativistic regime, plotted at each intersection of particle 1's coherent state components for $\sqrt{2\omega}\alpha=0.1$, and $\kappa=10^{-3}$. The detector is placed at $x = \sqrt{\frac{2}{\omega}} \alpha \cos\left(\frac{3\pi}{8}\right)$. The plot compares visibility with and without relativistic corrections $\Delta V_{rel}$. As entanglement builds, fringe visibility decreases, with relativistic effects accelerating this reduction.}
    \label{fig:visibility}
\end{figure}

For the final state considered here [Eq.(\ref{eqn:final state})], with $R$ on the $x$-axis when the two components of mass one are overlapping at $x=\sqrt{\frac{2}{\omega}}\alpha\cos{\pi/8}$, we have
\begin{equation}
    \begin{split}
        \expval{\phi^2_R}=|\mathcal{N}|^2|\psi_{\alpha_{1a}}(R)|^2\left(1+\mathrm{Re}\left(e^{i\sqrt{2\omega}\alpha R-\frac{\alpha^2}{\sqrt{2}}}\left(e^{i\left(\delta S^{(1)}_{aa}-\delta S^{(1)}_{ba}\right)}+e^{i\left(\delta S^{(1)}_{ab}-\delta S^{(1)}_{bb}\right)}\right)\right)\right).
    \end{split}
\end{equation}
This expression neglects the contributions from the probability of detecting one of the particles in the second mass which depend on $\psi_{\alpha_{2a}}$ and $\psi_{\alpha_{2b}}$, which are exponentially suppressed on the $x$-axis, as well as the terms proportional to $\braket{\alpha_{2a}}{\alpha_{2b}}$ which is also exponentially small. The fringe visibility is well approximated by
\begin{equation}
    V=1-\frac{1}{8}\left(\delta S^{(1)}_{aa}+\delta S^{(1)}_{bb}-\delta S^{(1)}_{ab}-\delta S^{(1)}_{ba}\right)^2,
\end{equation}
expanding in the gravitational coupling to lowest order. Using the forms for the $\delta S^{(1)}$ terms appropriate in the weakly relativistic regime [Eq. (\ref{eqn:ds1wr})], we can examine the corrections to the fringe visibility due to the lowest order relativistic corrections. We introduce the notation
\begin{equation}
    I^{NR}_{tot}=I^{NR}_{aa}+I^{NR}_{bb}-I^{NR}_{ba}-I^{NR}_{ba},
\end{equation}
and $I^{rel}_{tot}$ defined similarly, the fringe visibility in this regime (Fig. \ref{fig:visibility}) is then given by
\begin{equation}
\begin{split}
    V_{wr}=& 1-\frac{1}{8}\left(\frac{\kappa^2}{32\pi}\right)^2I^{NR}_{tot}
    \left(\omega t\right)\left(\left(\frac{1}{2\omega\alpha^2}+3\right)I^{NR}_{tot}\left(\omega t\right)+2I^{rel}_{tot}\left(\omega t\right)-\dot{I}_{tot}^{NR}(\omega t)\right).
\end{split}
\end{equation}
and the contribution from the relativistic effects can be written
\begin{equation}
\begin{split}
    \Delta V_{rel}=-&\frac{1}{8}\left(\frac{\kappa^2}{32\pi}\right)^2I^{NR}_{tot}\left(\omega t\right)\left(3I^{NR}_{tot}\left(\omega t\right)+2I^{rel}_{tot}\left(\omega t\right)-\dot{I}^{NR}_{tot}(\omega t)\right).
\end{split}
\end{equation}
The fringe visibility with and without the relativistic contribution is shown in Fig. \ref{fig:visibility}.

In this section, we derived the expected fringe visibility for our quantum field-theoretic model of gravitationally induced entanglement in the weakly relativistic regime. We showed that the dominant contribution to the relative phase evolution of the matter state arises from the perturbation to the classical action, which we expressed in terms of integrals capturing both nonrelativistic and relativistic effects. Expanding in the gravitational coupling, we obtained a correction to the fringe visibility due to relativistic effects, providing a direct means to quantify deviations from the nonrelativistic prediction. This showed how relativistic corrections accelerates the decay of fringe visibility in this model for gravitationally induced entanglement.

These results illustrate how relativistic corrections modify gravitationally induced entanglement within a quantum field-theoretic framework. While these effects remain small in the weakly relativistic regime, their inclusion provides a more complete description of the system’s evolution and could inform future analyses of relativistic corrections in quantum gravity experiments.

\section{\label{sec:conclusion}Conclusion}

In this work, we have explored the generation of entanglement between localized states of a scalar field due to their gravitational interaction. We developed an operational description of this entanglement using expectation values of field observables, allowing for a direct connection between theoretical predictions and measurable quantities. In our model, two massive objects—each in a superposition of localized position states—are described as excitations of a scalar field confined by an external harmonic potential. By formulating an operational approach to preparing coherent states of this field, we obtained an exact expression for a relativistic coherent state, which reduces to the familiar nonrelativistic coherent state in the appropriate limit. Numerical analysis revealed perturbations to the oscillation frequency and modulations of the variance in the weakly relativistic regime.

For the gravitational interaction, we derived an effective Hamiltonian by analyzing the time evolution operator for linearized quantum gravity in the static limit. This approach enabled us to isolate components of the evolution that entangle the matter fields with the gravitational field—terms associated with gravitational decoherence and backaction. By neglecting these terms, we traced out the gravitational degrees of freedom, leaving behind an effective self-interaction of the matter fields’ energy density. Applying this framework to our scalar field model led to a semiclassical approximation of the coherent state evolution under gravity.

The resulting dynamics exhibit entanglement between the field modes occupied by the relativistic coherent states. As the coherent state components of one of the masses evolve, they periodically overlap in position space, generating an interference pattern in the probability density of detecting one of them. This probability density is modeled by the expectation value of the field strength at the center of mass of the constituent particles of one of the masses. By analyzing this probability distribution, we calculated the fringe visibility and demonstrated that it decreases as gravitationally induced entanglement grows.

Importantly, this model allows us to incorporate relativistic effects in the gravitational interaction while remaining within a weakly relativistic framework. These effects introduce additional contributions to the fringe visibility, arising both from relativistic corrections to the initial energy of the states and from the coupling between amplitude and frequency in the relativistic harmonic trap, which serve to accelerate the decay of the fringe visibility.

Beyond this work, we plan to investigate gravitational decoherence in a similar setting \cite{blencowe_effective_2013}. Specifically, we will study how a superposition of coherent states in a harmonic trap coupled to linearized quantum gravity decoheres, leading to reductions in fringe visibility when the coherent state components overlap. This extends previous work on a zero-dimensional toy model of gravitational decoherence in a field-theoretic framework \cite{xu_zero-dimensional_2022}. In particular, we aim to explore the influence of non-vacuum graviton states—such as thermal or squeezed states—on the decoherence of the matter field, moving beyond the static limit.

Additionally, our results provide a foundation for further exploration of relativistic coherent states in quantum field theory. Future research will investigate the representation of the Heisenberg group that these states furnish on the Fock space of a quantum scalar field. In particular, we will analyze how the field's total momentum operator and its conjugate—which behaves as an effective position operator—form a representation of the canonical commutation relations. This conjugate operator, though central to localization in quantum field theory, does not transform as a vector under Lorentz transformations, making it an observer-dependent quantity. A deeper understanding of its properties, particularly in the context of many-particle subspaces, will help clarify its relationship to Newton-Wigner localization and other localization schemes \cite{barros_e_sa_quantum_2021}.

This work presents a quantum field-theoretic approach to gravitationally induced entanglement, demonstrating how relativistic effects modify the expected interference pattern in a weakly relativistic setting. By bridging the gap between quantum field-theoretic and operational descriptions of measurement, our model offers a new framework for understanding quantum coherence and gravitational interactions. These results not only contribute to the ongoing discussion of gravity's role in quantum mechanics but also lay the groundwork for further studies in relativistic quantum information, such as gravitational decoherence.

\section*{Acknowledgements}
We thank Charis Anastopoulos, Markus Aspelmeyer, Bei-Lok Hu, David Mattingly, Sougato Bose, Alexander Smith, Rufus Boyack, and Robert Caldwell for very helpful comments. This work was supported by the NSF under Grant No. PHY-2011382.

\bibliography{refs}

\begin{thebibliography}{51}%
\makeatletter
\providecommand \@ifxundefined [1]{%
 \@ifx{#1\undefined}
}%
\providecommand \@ifnum [1]{%
 \ifnum #1\expandafter \@firstoftwo
 \else \expandafter \@secondoftwo
 \fi
}%
\providecommand \@ifx [1]{%
 \ifx #1\expandafter \@firstoftwo
 \else \expandafter \@secondoftwo
 \fi
}%
\providecommand \natexlab [1]{#1}%
\providecommand \enquote  [1]{``#1''}%
\providecommand \bibnamefont  [1]{#1}%
\providecommand \bibfnamefont [1]{#1}%
\providecommand \citenamefont [1]{#1}%
\providecommand \href@noop [0]{\@secondoftwo}%
\providecommand \href [0]{\begingroup \@sanitize@url \@href}%
\providecommand \@href[1]{\@@startlink{#1}\@@href}%
\providecommand \@@href[1]{\endgroup#1\@@endlink}%
\providecommand \@sanitize@url [0]{\catcode `\\12\catcode `\$12\catcode `\&12\catcode `\#12\catcode `\^12\catcode `\_12\catcode `\%12\relax}%
\providecommand \@@startlink[1]{}%
\providecommand \@@endlink[0]{}%
\providecommand \url  [0]{\begingroup\@sanitize@url \@url }%
\providecommand \@url [1]{\endgroup\@href {#1}{\urlprefix }}%
\providecommand \urlprefix  [0]{URL }%
\providecommand \Eprint [0]{\href }%
\providecommand \doibase [0]{https://doi.org/}%
\providecommand \selectlanguage [0]{\@gobble}%
\providecommand \bibinfo  [0]{\@secondoftwo}%
\providecommand \bibfield  [0]{\@secondoftwo}%
\providecommand \translation [1]{[#1]}%
\providecommand \BibitemOpen [0]{}%
\providecommand \bibitemStop [0]{}%
\providecommand \bibitemNoStop [0]{.\EOS\space}%
\providecommand \EOS [0]{\spacefactor3000\relax}%
\providecommand \BibitemShut  [1]{\csname bibitem#1\endcsname}%
\let\auto@bib@innerbib\@empty
\bibitem [{\citenamefont {Macías}\ and\ \citenamefont {Camacho}(2008)}]{macias_incompatibility_2008}%
  \BibitemOpen
  \bibfield  {author} {\bibinfo {author} {\bibfnamefont {A.}~\bibnamefont {Macías}}\ and\ \bibinfo {author} {\bibfnamefont {A.}~\bibnamefont {Camacho}},\ }\bibfield  {title} {\bibinfo {title} {On the incompatibility between quantum theory and general relativity},\ }\href {https://doi.org/10.1016/j.physletb.2008.03.052} {\bibfield  {journal} {\bibinfo  {journal} {Physics Letters B}\ }\textbf {\bibinfo {volume} {663}},\ \bibinfo {pages} {99} (\bibinfo {year} {2008})}\BibitemShut {NoStop}%
\bibitem [{\citenamefont {Loll}\ \emph {et~al.}(2022)\citenamefont {Loll}, \citenamefont {Fabiano}, \citenamefont {Frattulillo},\ and\ \citenamefont {Wagner}}]{loll_quantum_2022}%
  \BibitemOpen
  \bibfield  {author} {\bibinfo {author} {\bibfnamefont {R.}~\bibnamefont {Loll}}, \bibinfo {author} {\bibfnamefont {G.}~\bibnamefont {Fabiano}}, \bibinfo {author} {\bibfnamefont {D.}~\bibnamefont {Frattulillo}},\ and\ \bibinfo {author} {\bibfnamefont {F.}~\bibnamefont {Wagner}},\ }\href {http://arxiv.org/abs/2206.06762} {\bibinfo {title} {Quantum {Gravity} in 30 {Questions}}} (\bibinfo {year} {2022}),\ \bibinfo {note} {arXiv:2206.06762 [hep-th]}\BibitemShut {NoStop}%
\bibitem [{\citenamefont {Carlip}(2001)}]{carlip_quantum_2001}%
  \BibitemOpen
  \bibfield  {author} {\bibinfo {author} {\bibfnamefont {S.}~\bibnamefont {Carlip}},\ }\bibfield  {title} {\bibinfo {title} {Quantum {Gravity}: a {Progress} {Report}},\ }\href {https://doi.org/10.1088/0034-4885/64/8/301} {\bibfield  {journal} {\bibinfo  {journal} {Reports on Progress in Physics}\ }\textbf {\bibinfo {volume} {64}},\ \bibinfo {pages} {885} (\bibinfo {year} {2001})},\ \bibinfo {note} {arXiv:gr-qc/0108040}\BibitemShut {NoStop}%
\bibitem [{\citenamefont {Kiefer}(2006)}]{kiefer_quantum_2006}%
  \BibitemOpen
  \bibfield  {author} {\bibinfo {author} {\bibfnamefont {C.}~\bibnamefont {Kiefer}},\ }\bibfield  {title} {\bibinfo {title} {Quantum {Gravity}: {General} {Introduction} and {Recent} {Developments}},\ }\href {https://doi.org/10.1002/andp.200510175} {\bibfield  {journal} {\bibinfo  {journal} {Annalen der Physik}\ }\textbf {\bibinfo {volume} {518}},\ \bibinfo {pages} {149} (\bibinfo {year} {2006})},\ \bibinfo {note} {arXiv:gr-qc/0508120}\BibitemShut {NoStop}%
\bibitem [{\citenamefont {Kiefer}(2013)}]{kiefer_conceptual_2013}%
  \BibitemOpen
  \bibfield  {author} {\bibinfo {author} {\bibfnamefont {C.}~\bibnamefont {Kiefer}},\ }\bibfield  {title} {\bibinfo {title} {Conceptual {Problems} in {Quantum} {Gravity} and {Quantum} {Cosmology}},\ }\href {https://doi.org/10.1155/2013/509316} {\bibfield  {journal} {\bibinfo  {journal} {ISRN Mathematical Physics}\ }\textbf {\bibinfo {volume} {2013}},\ \bibinfo {pages} {1} (\bibinfo {year} {2013})},\ \bibinfo {note} {arXiv:1401.3578 [gr-qc, physics:hep-th, physics:quant-ph]}\BibitemShut {NoStop}%
\bibitem [{\citenamefont {Oppenheim}(2023)}]{oppenheim_is_2023}%
  \BibitemOpen
  \bibfield  {author} {\bibinfo {author} {\bibfnamefont {J.}~\bibnamefont {Oppenheim}},\ }\href {http://arxiv.org/abs/2310.12221} {\bibinfo {title} {Is it time to rethink quantum gravity?}} (\bibinfo {year} {2023}),\ \bibinfo {note} {arXiv:2310.12221 [gr-qc, physics:hep-th, physics:quant-ph]}\BibitemShut {NoStop}%
\bibitem [{\citenamefont {Bose}\ \emph {et~al.}(2017)\citenamefont {Bose}, \citenamefont {Mazumdar}, \citenamefont {Morley}, \citenamefont {Ulbricht}, \citenamefont {Toroš}, \citenamefont {Paternostro}, \citenamefont {Geraci}, \citenamefont {Barker}, \citenamefont {Kim},\ and\ \citenamefont {Milburn}}]{bose_spin_2017}%
  \BibitemOpen
  \bibfield  {author} {\bibinfo {author} {\bibfnamefont {S.}~\bibnamefont {Bose}}, \bibinfo {author} {\bibfnamefont {A.}~\bibnamefont {Mazumdar}}, \bibinfo {author} {\bibfnamefont {G.~W.}\ \bibnamefont {Morley}}, \bibinfo {author} {\bibfnamefont {H.}~\bibnamefont {Ulbricht}}, \bibinfo {author} {\bibfnamefont {M.}~\bibnamefont {Toroš}}, \bibinfo {author} {\bibfnamefont {M.}~\bibnamefont {Paternostro}}, \bibinfo {author} {\bibfnamefont {A.~A.}\ \bibnamefont {Geraci}}, \bibinfo {author} {\bibfnamefont {P.~F.}\ \bibnamefont {Barker}}, \bibinfo {author} {\bibfnamefont {M.}~\bibnamefont {Kim}},\ and\ \bibinfo {author} {\bibfnamefont {G.}~\bibnamefont {Milburn}},\ }\bibfield  {title} {\bibinfo {title} {Spin {Entanglement} {Witness} for {Quantum} {Gravity}},\ }\href {https://doi.org/10.1103/PhysRevLett.119.240401} {\bibfield  {journal} {\bibinfo  {journal} {Physical Review Letters}\ }\textbf {\bibinfo {volume} {119}},\ \bibinfo {pages} {240401} (\bibinfo {year} {2017})}\BibitemShut {NoStop}%
\bibitem [{\citenamefont {Marletto}\ and\ \citenamefont {Vedral}(2017)}]{marletto_gravitationally_2017}%
  \BibitemOpen
  \bibfield  {author} {\bibinfo {author} {\bibfnamefont {C.}~\bibnamefont {Marletto}}\ and\ \bibinfo {author} {\bibfnamefont {V.}~\bibnamefont {Vedral}},\ }\bibfield  {title} {\bibinfo {title} {Gravitationally {Induced} {Entanglement} between {Two} {Massive} {Particles} is {Sufficient} {Evidence} of {Quantum} {Effects} in {Gravity}},\ }\href {https://doi.org/10.1103/PhysRevLett.119.240402} {\bibfield  {journal} {\bibinfo  {journal} {Physical Review Letters}\ }\textbf {\bibinfo {volume} {119}},\ \bibinfo {pages} {240402} (\bibinfo {year} {2017})}\BibitemShut {NoStop}%
\bibitem [{\citenamefont {Yi}\ \emph {et~al.}(2022)\citenamefont {Yi}, \citenamefont {Sinha}, \citenamefont {Home}, \citenamefont {Mazumdar},\ and\ \citenamefont {Bose}}]{yi_spatial_2022}%
  \BibitemOpen
  \bibfield  {author} {\bibinfo {author} {\bibfnamefont {B.}~\bibnamefont {Yi}}, \bibinfo {author} {\bibfnamefont {U.}~\bibnamefont {Sinha}}, \bibinfo {author} {\bibfnamefont {D.}~\bibnamefont {Home}}, \bibinfo {author} {\bibfnamefont {A.}~\bibnamefont {Mazumdar}},\ and\ \bibinfo {author} {\bibfnamefont {S.}~\bibnamefont {Bose}},\ }\href {http://arxiv.org/abs/2211.03661} {\bibinfo {title} {Spatial {Qubit} {Entanglement} {Witness} for {Quantum} {Natured} {Gravity}}} (\bibinfo {year} {2022}),\ \bibinfo {note} {arXiv:2211.03661 [gr-qc, physics:quant-ph]}\BibitemShut {NoStop}%
\bibitem [{\citenamefont {Carney}\ \emph {et~al.}(2021)\citenamefont {Carney}, \citenamefont {Müller},\ and\ \citenamefont {Taylor}}]{carney_using_2021}%
  \BibitemOpen
  \bibfield  {author} {\bibinfo {author} {\bibfnamefont {D.}~\bibnamefont {Carney}}, \bibinfo {author} {\bibfnamefont {H.}~\bibnamefont {Müller}},\ and\ \bibinfo {author} {\bibfnamefont {J.~M.}\ \bibnamefont {Taylor}},\ }\bibfield  {title} {\bibinfo {title} {Using an {Atom} {Interferometer} to {Infer} {Gravitational} {Entanglement} {Generation}},\ }\href {https://doi.org/10.1103/PRXQuantum.2.030330} {\bibfield  {journal} {\bibinfo  {journal} {PRX Quantum}\ }\textbf {\bibinfo {volume} {2}},\ \bibinfo {pages} {030330} (\bibinfo {year} {2021})}\BibitemShut {NoStop}%
\bibitem [{\citenamefont {Cui}\ and\ \citenamefont {Yi}(2023)}]{cui_exponentially_2023}%
  \BibitemOpen
  \bibfield  {author} {\bibinfo {author} {\bibfnamefont {D.}~\bibnamefont {Cui}}\ and\ \bibinfo {author} {\bibfnamefont {X.~X.}\ \bibnamefont {Yi}},\ }\bibfield  {title} {\bibinfo {title} {Exponentially enhanced gravitationally induced entanglement between quantum systems with a two-phonon drive},\ }\href {https://doi.org/10.1103/PhysRevA.108.023502} {\bibfield  {journal} {\bibinfo  {journal} {Physical Review A}\ }\textbf {\bibinfo {volume} {108}},\ \bibinfo {pages} {023502} (\bibinfo {year} {2023})},\ \bibinfo {note} {arXiv:2307.03657 [quant-ph]}\BibitemShut {NoStop}%
\bibitem [{\citenamefont {Feng}\ and\ \citenamefont {Vedral}(2022)}]{feng_amplification_2022}%
  \BibitemOpen
  \bibfield  {author} {\bibinfo {author} {\bibfnamefont {T.}~\bibnamefont {Feng}}\ and\ \bibinfo {author} {\bibfnamefont {V.}~\bibnamefont {Vedral}},\ }\bibfield  {title} {\bibinfo {title} {Amplification of {Gravitationally} {Induced} {Entanglement}},\ }\href {https://doi.org/10.1103/PhysRevD.106.066013} {\bibfield  {journal} {\bibinfo  {journal} {Physical Review D}\ }\textbf {\bibinfo {volume} {106}},\ \bibinfo {pages} {066013} (\bibinfo {year} {2022})},\ \bibinfo {note} {arXiv:2202.09737 [hep-th, physics:quant-ph]}\BibitemShut {NoStop}%
\bibitem [{\citenamefont {Guff}\ \emph {et~al.}(2022)\citenamefont {Guff}, \citenamefont {Boulle},\ and\ \citenamefont {Pikovski}}]{guff_optimal_2022}%
  \BibitemOpen
  \bibfield  {author} {\bibinfo {author} {\bibfnamefont {T.}~\bibnamefont {Guff}}, \bibinfo {author} {\bibfnamefont {N.}~\bibnamefont {Boulle}},\ and\ \bibinfo {author} {\bibfnamefont {I.}~\bibnamefont {Pikovski}},\ }\bibfield  {title} {\bibinfo {title} {Optimal fidelity witnesses for gravitational entanglement},\ }\href {https://doi.org/10.1103/PhysRevA.105.022444} {\bibfield  {journal} {\bibinfo  {journal} {Physical Review A}\ }\textbf {\bibinfo {volume} {105}},\ \bibinfo {pages} {022444} (\bibinfo {year} {2022})}\BibitemShut {NoStop}%
\bibitem [{\citenamefont {Tilly}\ \emph {et~al.}(2021)\citenamefont {Tilly}, \citenamefont {Marshman}, \citenamefont {Mazumdar},\ and\ \citenamefont {Bose}}]{tilly_qudits_2021}%
  \BibitemOpen
  \bibfield  {author} {\bibinfo {author} {\bibfnamefont {J.}~\bibnamefont {Tilly}}, \bibinfo {author} {\bibfnamefont {R.~J.}\ \bibnamefont {Marshman}}, \bibinfo {author} {\bibfnamefont {A.}~\bibnamefont {Mazumdar}},\ and\ \bibinfo {author} {\bibfnamefont {S.}~\bibnamefont {Bose}},\ }\bibfield  {title} {\bibinfo {title} {Qudits for witnessing quantum-gravity-induced entanglement of masses under decoherence},\ }\href {https://doi.org/10.1103/PhysRevA.104.052416} {\bibfield  {journal} {\bibinfo  {journal} {Physical Review A}\ }\textbf {\bibinfo {volume} {104}},\ \bibinfo {pages} {052416} (\bibinfo {year} {2021})}\BibitemShut {NoStop}%
\bibitem [{\citenamefont {Fujita}\ \emph {et~al.}(2023)\citenamefont {Fujita}, \citenamefont {Kaku}, \citenamefont {Matumura},\ and\ \citenamefont {Michimura}}]{fujita_inverted_2023}%
  \BibitemOpen
  \bibfield  {author} {\bibinfo {author} {\bibfnamefont {T.}~\bibnamefont {Fujita}}, \bibinfo {author} {\bibfnamefont {Y.}~\bibnamefont {Kaku}}, \bibinfo {author} {\bibfnamefont {A.}~\bibnamefont {Matumura}},\ and\ \bibinfo {author} {\bibfnamefont {Y.}~\bibnamefont {Michimura}},\ }\href {http://arxiv.org/abs/2308.14552} {\bibinfo {title} {Inverted {Oscillators} for {Testing} {Gravity}-induced {Quantum} {Entanglement}}} (\bibinfo {year} {2023}),\ \bibinfo {note} {arXiv:2308.14552 [gr-qc, physics:quant-ph]}\BibitemShut {NoStop}%
\bibitem [{\citenamefont {Kaku}\ \emph {et~al.}(2025)\citenamefont {Kaku}, \citenamefont {Matsumura},\ and\ \citenamefont {Fujita}}]{kaku_sudden_2025}%
  \BibitemOpen
  \bibfield  {author} {\bibinfo {author} {\bibfnamefont {Y.}~\bibnamefont {Kaku}}, \bibinfo {author} {\bibfnamefont {A.}~\bibnamefont {Matsumura}},\ and\ \bibinfo {author} {\bibfnamefont {T.}~\bibnamefont {Fujita}},\ }\href {https://doi.org/10.48550/arXiv.2501.18147} {\bibinfo {title} {Sudden {Decoherence} by {Resonant} {Particle} {Excitation} for {Testing} {Gravity}-{Induced} {Entanglement}}} (\bibinfo {year} {2025}),\ \bibinfo {note} {arXiv:2501.18147 [quant-ph]}\BibitemShut {NoStop}%
\bibitem [{\citenamefont {Miki}\ \emph {et~al.}(2023)\citenamefont {Miki}, \citenamefont {Matsumura},\ and\ \citenamefont {Yamamoto}}]{miki_quantum_2023}%
  \BibitemOpen
  \bibfield  {author} {\bibinfo {author} {\bibfnamefont {D.}~\bibnamefont {Miki}}, \bibinfo {author} {\bibfnamefont {A.}~\bibnamefont {Matsumura}},\ and\ \bibinfo {author} {\bibfnamefont {K.}~\bibnamefont {Yamamoto}},\ }\href {http://arxiv.org/abs/2311.00563} {\bibinfo {title} {Quantum signature of gravity in optomechanical systems with conditional measurement}} (\bibinfo {year} {2023}),\ \bibinfo {note} {arXiv:2311.00563 [gr-qc, physics:quant-ph]}\BibitemShut {NoStop}%
\bibitem [{\citenamefont {Schut}\ \emph {et~al.}(2023)\citenamefont {Schut}, \citenamefont {Grinin}, \citenamefont {Dana}, \citenamefont {Bose}, \citenamefont {Geraci},\ and\ \citenamefont {Mazumdar}}]{schut_relaxation_2023}%
  \BibitemOpen
  \bibfield  {author} {\bibinfo {author} {\bibfnamefont {M.}~\bibnamefont {Schut}}, \bibinfo {author} {\bibfnamefont {A.}~\bibnamefont {Grinin}}, \bibinfo {author} {\bibfnamefont {A.}~\bibnamefont {Dana}}, \bibinfo {author} {\bibfnamefont {S.}~\bibnamefont {Bose}}, \bibinfo {author} {\bibfnamefont {A.}~\bibnamefont {Geraci}},\ and\ \bibinfo {author} {\bibfnamefont {A.}~\bibnamefont {Mazumdar}},\ }\href {http://arxiv.org/abs/2307.07536} {\bibinfo {title} {Relaxation of experimental parameters in a {Quantum}-{Gravity} {Induced} {Entanglement} of {Masses} {Protocol} using electromagnetic screening}} (\bibinfo {year} {2023}),\ \bibinfo {note} {arXiv:2307.07536 [gr-qc, physics:quant-ph]}\BibitemShut {NoStop}%
\bibitem [{\citenamefont {Patrascu}(2023)}]{patrascu_graviton_2023}%
  \BibitemOpen
  \bibfield  {author} {\bibinfo {author} {\bibfnamefont {A.~T.}\ \bibnamefont {Patrascu}},\ }\href {http://arxiv.org/abs/2311.10715} {\bibinfo {title} {Graviton mediated polarisation-polarisation entanglement of photons by means of the {Schwinger} {Keldysh} and {Kadanoff} {Baym} formalisms and {Quantum} {Boltzmann} equations}} (\bibinfo {year} {2023}),\ \bibinfo {note} {arXiv:2311.10715 [hep-th]}\BibitemShut {NoStop}%
\bibitem [{\citenamefont {Anastopoulos}\ and\ \citenamefont {Hu}(2022)}]{anastopoulos_gravity_2022}%
  \BibitemOpen
  \bibfield  {author} {\bibinfo {author} {\bibfnamefont {C.}~\bibnamefont {Anastopoulos}}\ and\ \bibinfo {author} {\bibfnamefont {B.-L.}\ \bibnamefont {Hu}},\ }\bibfield  {title} {\bibinfo {title} {Gravity, {Quantum} {Fields} and {Quantum} {Information}: {Problems} with classical channel and stochastic theories},\ }\href {https://doi.org/10.3390/e24040490} {\bibfield  {journal} {\bibinfo  {journal} {Entropy}\ }\textbf {\bibinfo {volume} {24}},\ \bibinfo {pages} {490} (\bibinfo {year} {2022})},\ \bibinfo {note} {arXiv:2202.02789 [gr-qc, physics:quant-ph]}\BibitemShut {NoStop}%
\bibitem [{\citenamefont {Christodoulou}\ \emph {et~al.}(2023{\natexlab{a}})\citenamefont {Christodoulou}, \citenamefont {Di~Biagio}, \citenamefont {Aspelmeyer}, \citenamefont {Brukner}, \citenamefont {Rovelli},\ and\ \citenamefont {Howl}}]{christodoulou_locally_2023}%
  \BibitemOpen
  \bibfield  {author} {\bibinfo {author} {\bibfnamefont {M.}~\bibnamefont {Christodoulou}}, \bibinfo {author} {\bibfnamefont {A.}~\bibnamefont {Di~Biagio}}, \bibinfo {author} {\bibfnamefont {M.}~\bibnamefont {Aspelmeyer}}, \bibinfo {author} {\bibfnamefont {{\v{C}}.}~\bibnamefont {Brukner}}, \bibinfo {author} {\bibfnamefont {C.}~\bibnamefont {Rovelli}},\ and\ \bibinfo {author} {\bibfnamefont {R.}~\bibnamefont {Howl}},\ }\bibfield  {title} {\bibinfo {title} {Locally mediated entanglement in linearised quantum gravity},\ }\href {https://doi.org/10.1103/PhysRevLett.130.100202} {\bibfield  {journal} {\bibinfo  {journal} {Physical Review Letters}\ }\textbf {\bibinfo {volume} {130}},\ \bibinfo {pages} {100202} (\bibinfo {year} {2023}{\natexlab{a}})},\ \bibinfo {note} {arXiv:2202.03368 [gr-qc, physics:quant-ph]}\BibitemShut {NoStop}%
\bibitem [{\citenamefont {Christodoulou}\ \emph {et~al.}(2023{\natexlab{b}})\citenamefont {Christodoulou}, \citenamefont {Di~Biagio}, \citenamefont {Howl},\ and\ \citenamefont {Rovelli}}]{christodoulou_gravity_2023}%
  \BibitemOpen
  \bibfield  {author} {\bibinfo {author} {\bibfnamefont {M.}~\bibnamefont {Christodoulou}}, \bibinfo {author} {\bibfnamefont {A.}~\bibnamefont {Di~Biagio}}, \bibinfo {author} {\bibfnamefont {R.}~\bibnamefont {Howl}},\ and\ \bibinfo {author} {\bibfnamefont {C.}~\bibnamefont {Rovelli}},\ }\bibfield  {title} {\bibinfo {title} {Gravity entanglement, quantum reference systems, degrees of freedom},\ }\href {https://doi.org/10.1088/1361-6382/acb0aa} {\bibfield  {journal} {\bibinfo  {journal} {Classical and Quantum Gravity}\ }\textbf {\bibinfo {volume} {40}},\ \bibinfo {pages} {047001} (\bibinfo {year} {2023}{\natexlab{b}})},\ \bibinfo {note} {arXiv:2207.03138 [gr-qc, physics:quant-ph]}\BibitemShut {NoStop}%
\bibitem [{\citenamefont {Bose}\ \emph {et~al.}(2022)\citenamefont {Bose}, \citenamefont {Mazumdar}, \citenamefont {Schut},\ and\ \citenamefont {Toroš}}]{bose_mechanism_2022}%
  \BibitemOpen
  \bibfield  {author} {\bibinfo {author} {\bibfnamefont {S.}~\bibnamefont {Bose}}, \bibinfo {author} {\bibfnamefont {A.}~\bibnamefont {Mazumdar}}, \bibinfo {author} {\bibfnamefont {M.}~\bibnamefont {Schut}},\ and\ \bibinfo {author} {\bibfnamefont {M.}~\bibnamefont {Toroš}},\ }\bibfield  {title} {\bibinfo {title} {Mechanism for the quantum natured gravitons to entangle masses},\ }\href {https://doi.org/10.1103/PhysRevD.105.106028} {\bibfield  {journal} {\bibinfo  {journal} {Physical Review D}\ }\textbf {\bibinfo {volume} {105}},\ \bibinfo {pages} {106028} (\bibinfo {year} {2022})},\ \bibinfo {note} {arXiv:2201.03583 [gr-qc, physics:hep-th, physics:quant-ph]}\BibitemShut {NoStop}%
\bibitem [{\citenamefont {Danielson}\ \emph {et~al.}(2022)\citenamefont {Danielson}, \citenamefont {Satishchandran},\ and\ \citenamefont {Wald}}]{danielson_gravitationally_2022}%
  \BibitemOpen
  \bibfield  {author} {\bibinfo {author} {\bibfnamefont {D.~L.}\ \bibnamefont {Danielson}}, \bibinfo {author} {\bibfnamefont {G.}~\bibnamefont {Satishchandran}},\ and\ \bibinfo {author} {\bibfnamefont {R.~M.}\ \bibnamefont {Wald}},\ }\bibfield  {title} {\bibinfo {title} {Gravitationally mediated entanglement: {Newtonian} field versus gravitons},\ }\href {https://doi.org/10.1103/PhysRevD.105.086001} {\bibfield  {journal} {\bibinfo  {journal} {Physical Review D}\ }\textbf {\bibinfo {volume} {105}},\ \bibinfo {pages} {086001} (\bibinfo {year} {2022})}\BibitemShut {NoStop}%
\bibitem [{\citenamefont {Ma}\ \emph {et~al.}(2022)\citenamefont {Ma}, \citenamefont {Guff}, \citenamefont {Morley}, \citenamefont {Pikovski},\ and\ \citenamefont {Kim}}]{ma_limits_2022}%
  \BibitemOpen
  \bibfield  {author} {\bibinfo {author} {\bibfnamefont {Y.}~\bibnamefont {Ma}}, \bibinfo {author} {\bibfnamefont {T.}~\bibnamefont {Guff}}, \bibinfo {author} {\bibfnamefont {G.}~\bibnamefont {Morley}}, \bibinfo {author} {\bibfnamefont {I.}~\bibnamefont {Pikovski}},\ and\ \bibinfo {author} {\bibfnamefont {M.~S.}\ \bibnamefont {Kim}},\ }\bibfield  {title} {\bibinfo {title} {Limits on inference of gravitational entanglement},\ }\href {https://doi.org/10.1103/PhysRevResearch.4.013024} {\bibfield  {journal} {\bibinfo  {journal} {Physical Review Research}\ }\textbf {\bibinfo {volume} {4}},\ \bibinfo {pages} {013024} (\bibinfo {year} {2022})},\ \bibinfo {note} {arXiv:2111.00936 [quant-ph]}\BibitemShut {NoStop}%
\bibitem [{\citenamefont {Marletto}\ and\ \citenamefont {Vedral}(2018)}]{marletto_when_2018}%
  \BibitemOpen
  \bibfield  {author} {\bibinfo {author} {\bibfnamefont {C.}~\bibnamefont {Marletto}}\ and\ \bibinfo {author} {\bibfnamefont {V.}~\bibnamefont {Vedral}},\ }\bibfield  {title} {\bibinfo {title} {When can gravity path-entangle two spatially superposed masses?},\ }\href {https://doi.org/10.1103/PhysRevD.98.046001} {\bibfield  {journal} {\bibinfo  {journal} {Physical Review D}\ }\textbf {\bibinfo {volume} {98}},\ \bibinfo {pages} {046001} (\bibinfo {year} {2018})},\ \bibinfo {note} {arXiv:1803.09124 [quant-ph]}\BibitemShut {NoStop}%
\bibitem [{\citenamefont {Martín-Martínez}\ and\ \citenamefont {Perche}(2023)}]{martin-martinez_what_2023}%
  \BibitemOpen
  \bibfield  {author} {\bibinfo {author} {\bibfnamefont {E.}~\bibnamefont {Martín-Martínez}}\ and\ \bibinfo {author} {\bibfnamefont {T.~R.}\ \bibnamefont {Perche}},\ }\bibfield  {title} {\bibinfo {title} {What gravity mediated entanglement can really tell us about quantum gravity},\ }\href {https://doi.org/10.1103/PhysRevD.108.L101702} {\bibfield  {journal} {\bibinfo  {journal} {Physical Review D}\ }\textbf {\bibinfo {volume} {108}},\ \bibinfo {pages} {L101702} (\bibinfo {year} {2023})},\ \bibinfo {note} {arXiv:2208.09489 [gr-qc, physics:hep-th, physics:quant-ph]}\BibitemShut {NoStop}%
\bibitem [{\citenamefont {Rydving}\ \emph {et~al.}(2021)\citenamefont {Rydving}, \citenamefont {Aurell},\ and\ \citenamefont {Pikovski}}]{rydving_gedanken_2021}%
  \BibitemOpen
  \bibfield  {author} {\bibinfo {author} {\bibfnamefont {E.}~\bibnamefont {Rydving}}, \bibinfo {author} {\bibfnamefont {E.}~\bibnamefont {Aurell}},\ and\ \bibinfo {author} {\bibfnamefont {I.}~\bibnamefont {Pikovski}},\ }\bibfield  {title} {\bibinfo {title} {Do {Gedanken} experiments compel quantization of gravity?},\ }\href {https://doi.org/10.1103/PhysRevD.104.086024} {\bibfield  {journal} {\bibinfo  {journal} {Physical Review D}\ }\textbf {\bibinfo {volume} {104}},\ \bibinfo {pages} {086024} (\bibinfo {year} {2021})}\BibitemShut {NoStop}%
\bibitem [{\citenamefont {Di~Biagio}\ \emph {et~al.}(2023)\citenamefont {Di~Biagio}, \citenamefont {Howl}, \citenamefont {Brukner}, \citenamefont {Rovelli},\ and\ \citenamefont {Christodoulou}}]{di_biagio_relativistic_2023}%
  \BibitemOpen
  \bibfield  {author} {\bibinfo {author} {\bibfnamefont {A.}~\bibnamefont {Di~Biagio}}, \bibinfo {author} {\bibfnamefont {R.}~\bibnamefont {Howl}}, \bibinfo {author} {\bibfnamefont {{\v{C}}.}~\bibnamefont {Brukner}}, \bibinfo {author} {\bibfnamefont {C.}~\bibnamefont {Rovelli}},\ and\ \bibinfo {author} {\bibfnamefont {M.}~\bibnamefont {Christodoulou}},\ }\href {http://arxiv.org/abs/2305.05645} {\bibinfo {title} {Relativistic locality can imply subsystem locality}} (\bibinfo {year} {2023}),\ \bibinfo {note} {arXiv:2305.05645 [quant-ph]}\BibitemShut {NoStop}%
\bibitem [{\citenamefont {Carney}(2022)}]{carney_newton_2022}%
  \BibitemOpen
  \bibfield  {author} {\bibinfo {author} {\bibfnamefont {D.}~\bibnamefont {Carney}},\ }\bibfield  {title} {\bibinfo {title} {Newton, entanglement, and the graviton},\ }\href {https://doi.org/10.1103/PhysRevD.105.024029} {\bibfield  {journal} {\bibinfo  {journal} {Physical Review D}\ }\textbf {\bibinfo {volume} {105}},\ \bibinfo {pages} {024029} (\bibinfo {year} {2022})},\ \bibinfo {note} {arXiv:2108.06320 [gr-qc, physics:hep-ph, physics:hep-th, physics:quant-ph]}\BibitemShut {NoStop}%
\bibitem [{\citenamefont {Chen}\ and\ \citenamefont {Giacomini}(2024)}]{chen_quantum_2024}%
  \BibitemOpen
  \bibfield  {author} {\bibinfo {author} {\bibfnamefont {L.-Q.}\ \bibnamefont {Chen}}\ and\ \bibinfo {author} {\bibfnamefont {F.}~\bibnamefont {Giacomini}},\ }\href {http://arxiv.org/abs/2402.10288} {\bibinfo {title} {Quantum effects in gravity beyond the {Newton} potential from a delocalised quantum source}} (\bibinfo {year} {2024}),\ \bibinfo {note} {arXiv:2402.10288 [quant-ph]}\BibitemShut {NoStop}%
\bibitem [{\citenamefont {Higgins}\ \emph {et~al.}(2024)\citenamefont {Higgins}, \citenamefont {Di~Biagio},\ and\ \citenamefont {Christodoulou}}]{higgins_gravitationally_2024}%
  \BibitemOpen
  \bibfield  {author} {\bibinfo {author} {\bibfnamefont {G.}~\bibnamefont {Higgins}}, \bibinfo {author} {\bibfnamefont {A.}~\bibnamefont {Di~Biagio}},\ and\ \bibinfo {author} {\bibfnamefont {M.}~\bibnamefont {Christodoulou}},\ }\href {http://arxiv.org/abs/2403.02062} {\bibinfo {title} {Gravitationally {Mediated} {Entanglement} with {Superpositions} of {Rotational} {Energies}}} (\bibinfo {year} {2024}),\ \bibinfo {note} {arXiv:2403.02062 [gr-qc, physics:quant-ph]}\BibitemShut {NoStop}%
\bibitem [{\citenamefont {Kaku}\ and\ \citenamefont {Nambu}(2024)}]{kaku_gravitational_2024}%
  \BibitemOpen
  \bibfield  {author} {\bibinfo {author} {\bibfnamefont {Y.}~\bibnamefont {Kaku}}\ and\ \bibinfo {author} {\bibfnamefont {Y.}~\bibnamefont {Nambu}},\ }\href {http://arxiv.org/abs/2411.12997} {\bibinfo {title} {Gravitational entanglement witness through {Einstein} ring image}} (\bibinfo {year} {2024}),\ \bibinfo {note} {arXiv:2411.12997 [gr-qc]}\BibitemShut {NoStop}%
\bibitem [{\citenamefont {Tobar}\ \emph {et~al.}(2023)\citenamefont {Tobar}, \citenamefont {Manikandan}, \citenamefont {Beitel},\ and\ \citenamefont {Pikovski}}]{tobar_detecting_2023}%
  \BibitemOpen
  \bibfield  {author} {\bibinfo {author} {\bibfnamefont {G.}~\bibnamefont {Tobar}}, \bibinfo {author} {\bibfnamefont {S.~K.}\ \bibnamefont {Manikandan}}, \bibinfo {author} {\bibfnamefont {T.}~\bibnamefont {Beitel}},\ and\ \bibinfo {author} {\bibfnamefont {I.}~\bibnamefont {Pikovski}},\ }\href {http://arxiv.org/abs/2308.15440} {\bibinfo {title} {Detecting single gravitons with quantum sensing}} (\bibinfo {year} {2023}),\ \bibinfo {note} {arXiv:2308.15440 [astro-ph, physics:gr-qc, physics:hep-th, physics:quant-ph]}\BibitemShut {NoStop}%
\bibitem [{\citenamefont {Bose}\ \emph {et~al.}(2023)\citenamefont {Bose}, \citenamefont {Fuentes}, \citenamefont {Geraci}, \citenamefont {Khan}, \citenamefont {Qvarfort}, \citenamefont {Rademacher}, \citenamefont {Rashid}, \citenamefont {Toroš}, \citenamefont {Ulbricht},\ and\ \citenamefont {Wanjura}}]{bose_massive_2023}%
  \BibitemOpen
  \bibfield  {author} {\bibinfo {author} {\bibfnamefont {S.}~\bibnamefont {Bose}}, \bibinfo {author} {\bibfnamefont {I.}~\bibnamefont {Fuentes}}, \bibinfo {author} {\bibfnamefont {A.~A.}\ \bibnamefont {Geraci}}, \bibinfo {author} {\bibfnamefont {S.~M.}\ \bibnamefont {Khan}}, \bibinfo {author} {\bibfnamefont {S.}~\bibnamefont {Qvarfort}}, \bibinfo {author} {\bibfnamefont {M.}~\bibnamefont {Rademacher}}, \bibinfo {author} {\bibfnamefont {M.}~\bibnamefont {Rashid}}, \bibinfo {author} {\bibfnamefont {M.}~\bibnamefont {Toroš}}, \bibinfo {author} {\bibfnamefont {H.}~\bibnamefont {Ulbricht}},\ and\ \bibinfo {author} {\bibfnamefont {C.~C.}\ \bibnamefont {Wanjura}},\ }\href {http://arxiv.org/abs/2311.09218} {\bibinfo {title} {Massive quantum systems as interfaces of quantum mechanics and gravity}} (\bibinfo {year} {2023}),\ \bibinfo {note} {arXiv:2311.09218 [quant-ph]}\BibitemShut {NoStop}%
\bibitem [{\citenamefont {Howl}\ \emph {et~al.}(2021)\citenamefont {Howl}, \citenamefont {Vedral}, \citenamefont {Naik}, \citenamefont {Christodoulou}, \citenamefont {Rovelli},\ and\ \citenamefont {Iyer}}]{howl_non-gaussianity_2021}%
  \BibitemOpen
  \bibfield  {author} {\bibinfo {author} {\bibfnamefont {R.}~\bibnamefont {Howl}}, \bibinfo {author} {\bibfnamefont {V.}~\bibnamefont {Vedral}}, \bibinfo {author} {\bibfnamefont {D.}~\bibnamefont {Naik}}, \bibinfo {author} {\bibfnamefont {M.}~\bibnamefont {Christodoulou}}, \bibinfo {author} {\bibfnamefont {C.}~\bibnamefont {Rovelli}},\ and\ \bibinfo {author} {\bibfnamefont {A.}~\bibnamefont {Iyer}},\ }\bibfield  {title} {\bibinfo {title} {Non-{Gaussianity} as a {Signature} of a {Quantum} {Theory} of {Gravity}},\ }\href {https://doi.org/10.1103/PRXQuantum.2.010325} {\bibfield  {journal} {\bibinfo  {journal} {PRX Quantum}\ }\textbf {\bibinfo {volume} {2}},\ \bibinfo {pages} {010325} (\bibinfo {year} {2021})}\BibitemShut {NoStop}%
\bibitem [{\citenamefont {Maleki}\ and\ \citenamefont {Maleki}(2022)}]{maleki_complementarity-entanglement_2022}%
  \BibitemOpen
  \bibfield  {author} {\bibinfo {author} {\bibfnamefont {Y.}~\bibnamefont {Maleki}}\ and\ \bibinfo {author} {\bibfnamefont {A.}~\bibnamefont {Maleki}},\ }\bibfield  {title} {\bibinfo {title} {Complementarity-{Entanglement} {Tradeoff} in {Quantum} {Gravity}},\ }\href {https://doi.org/10.1103/PhysRevD.105.086024} {\bibfield  {journal} {\bibinfo  {journal} {Physical Review D}\ }\textbf {\bibinfo {volume} {105}},\ \bibinfo {pages} {086024} (\bibinfo {year} {2022})},\ \bibinfo {note} {arXiv:2205.01967 [gr-qc, physics:hep-th, physics:quant-ph]}\BibitemShut {NoStop}%
\bibitem [{\citenamefont {Carney}\ \emph {et~al.}(2025)\citenamefont {Carney}, \citenamefont {Karydas}, \citenamefont {Scharnhorst}, \citenamefont {Singh},\ and\ \citenamefont {Taylor}}]{carney_quantum_2025}%
  \BibitemOpen
  \bibfield  {author} {\bibinfo {author} {\bibfnamefont {D.}~\bibnamefont {Carney}}, \bibinfo {author} {\bibfnamefont {M.}~\bibnamefont {Karydas}}, \bibinfo {author} {\bibfnamefont {T.}~\bibnamefont {Scharnhorst}}, \bibinfo {author} {\bibfnamefont {R.}~\bibnamefont {Singh}},\ and\ \bibinfo {author} {\bibfnamefont {J.~M.}\ \bibnamefont {Taylor}},\ }\href {https://doi.org/10.48550/arXiv.2502.17575} {\bibinfo {title} {On the quantum mechanics of entropic forces}} (\bibinfo {year} {2025}),\ \bibinfo {note} {arXiv:2502.17575 [hep-th]}\BibitemShut {NoStop}%
\bibitem [{\citenamefont {Kryhin}\ and\ \citenamefont {Sudhir}(2023)}]{kryhin_distinguishable_2023}%
  \BibitemOpen
  \bibfield  {author} {\bibinfo {author} {\bibfnamefont {S.}~\bibnamefont {Kryhin}}\ and\ \bibinfo {author} {\bibfnamefont {V.}~\bibnamefont {Sudhir}},\ }\href {http://arxiv.org/abs/2309.09105} {\bibinfo {title} {Distinguishable consequence of classical gravity on quantum matter}} (\bibinfo {year} {2023}),\ \bibinfo {note} {arXiv:2309.09105 [gr-qc, physics:hep-th, physics:quant-ph]}\BibitemShut {NoStop}%
\bibitem [{\citenamefont {Donoghue}(2012)}]{donoghue_effective_2012}%
  \BibitemOpen
  \bibfield  {author} {\bibinfo {author} {\bibfnamefont {J.~F.}\ \bibnamefont {Donoghue}},\ }\bibfield  {title} {\bibinfo {title} {The effective field theory treatment of quantum gravity}\ }(\bibinfo {year} {2012})\ pp.\ \bibinfo {pages} {73--94},\ \bibinfo {note} {arXiv:1209.3511 [gr-qc, physics:hep-ph, physics:hep-th]}\BibitemShut {NoStop}%
\bibitem [{\citenamefont {Heller}(1975)}]{heller_time-dependent_1975}%
  \BibitemOpen
  \bibfield  {author} {\bibinfo {author} {\bibfnamefont {E.~J.}\ \bibnamefont {Heller}},\ }\bibfield  {title} {\bibinfo {title} {Time-dependent approach to semiclassical dynamics},\ }\href {https://doi.org/10.1063/1.430620} {\bibfield  {journal} {\bibinfo  {journal} {The Journal of Chemical Physics}\ }\textbf {\bibinfo {volume} {62}},\ \bibinfo {pages} {1544} (\bibinfo {year} {1975})}\BibitemShut {NoStop}%
\bibitem [{\citenamefont {Yant}\ and\ \citenamefont {Blencowe}(2023)}]{yant_gravitationally_2023}%
  \BibitemOpen
  \bibfield  {author} {\bibinfo {author} {\bibfnamefont {J.}~\bibnamefont {Yant}}\ and\ \bibinfo {author} {\bibfnamefont {M.}~\bibnamefont {Blencowe}},\ }\bibfield  {title} {\bibinfo {title} {Gravitationally induced entanglement in a harmonic trap},\ }\href {https://doi.org/10.1103/PhysRevD.107.106018} {\bibfield  {journal} {\bibinfo  {journal} {Physical Review D}\ }\textbf {\bibinfo {volume} {107}},\ \bibinfo {pages} {106018} (\bibinfo {year} {2023})}\BibitemShut {NoStop}%
\bibitem [{\citenamefont {Oniga}\ and\ \citenamefont {Wang}(2017)}]{oniga_quantum_2017}%
  \BibitemOpen
  \bibfield  {author} {\bibinfo {author} {\bibfnamefont {T.}~\bibnamefont {Oniga}}\ and\ \bibinfo {author} {\bibfnamefont {C.~H.-T.}\ \bibnamefont {Wang}},\ }\bibfield  {title} {\bibinfo {title} {Quantum coherence, radiance, and resistance of gravitational systems},\ }\href {https://doi.org/10.1103/PhysRevD.96.084014} {\bibfield  {journal} {\bibinfo  {journal} {Physical Review D}\ }\textbf {\bibinfo {volume} {96}},\ \bibinfo {pages} {084014} (\bibinfo {year} {2017})}\BibitemShut {NoStop}%
\bibitem [{\citenamefont {Perche}\ \emph {et~al.}(2024)\citenamefont {Perche}, \citenamefont {Polo-Gómez}, \citenamefont {Torres},\ and\ \citenamefont {Martín-Martínez}}]{perche_particle_2024}%
  \BibitemOpen
  \bibfield  {author} {\bibinfo {author} {\bibfnamefont {T.~R.}\ \bibnamefont {Perche}}, \bibinfo {author} {\bibfnamefont {J.}~\bibnamefont {Polo-Gómez}}, \bibinfo {author} {\bibfnamefont {B.~d. S.~L.}\ \bibnamefont {Torres}},\ and\ \bibinfo {author} {\bibfnamefont {E.}~\bibnamefont {Martín-Martínez}},\ }\bibfield  {title} {\bibinfo {title} {Particle {Detectors} from {Localized} {Quantum} {Field} {Theories}},\ }\href {https://doi.org/10.1103/PhysRevD.109.045013} {\bibfield  {journal} {\bibinfo  {journal} {Physical Review D}\ }\textbf {\bibinfo {volume} {109}},\ \bibinfo {pages} {045013} (\bibinfo {year} {2024})},\ \bibinfo {note} {arXiv:2308.11698 [gr-qc, physics:hep-th, physics:quant-ph]}\BibitemShut {NoStop}%
\bibitem [{\citenamefont {Torres}(2024)}]{torres_particle_2024}%
  \BibitemOpen
  \bibfield  {author} {\bibinfo {author} {\bibfnamefont {B.~d. S.~L.}\ \bibnamefont {Torres}},\ }\bibfield  {title} {\bibinfo {title} {Particle detector models from path integrals of localized quantum fields},\ }\href {https://doi.org/10.1103/PhysRevD.109.065004} {\bibfield  {journal} {\bibinfo  {journal} {Physical Review D}\ }\textbf {\bibinfo {volume} {109}},\ \bibinfo {pages} {065004} (\bibinfo {year} {2024})},\ \bibinfo {note} {arXiv:2310.16083 [gr-qc, physics:hep-th, physics:quant-ph]}\BibitemShut {NoStop}%
\bibitem [{\citenamefont {Ragula}\ \emph {et~al.}(2025)\citenamefont {Ragula}, \citenamefont {Torres}, \citenamefont {Schnetter},\ and\ \citenamefont {Martín-Martínez}}]{ragula_localizing_2025}%
  \BibitemOpen
  \bibfield  {author} {\bibinfo {author} {\bibfnamefont {B.}~\bibnamefont {Ragula}}, \bibinfo {author} {\bibfnamefont {B.~d. S.~L.}\ \bibnamefont {Torres}}, \bibinfo {author} {\bibfnamefont {E.}~\bibnamefont {Schnetter}},\ and\ \bibinfo {author} {\bibfnamefont {E.}~\bibnamefont {Martín-Martínez}},\ }\href {https://doi.org/10.48550/arXiv.2502.02643} {\bibinfo {title} {Localizing quantum fields with time-dependent potentials}} (\bibinfo {year} {2025}),\ \bibinfo {note} {arXiv:2502.02643 [quant-ph]}\BibitemShut {NoStop}%
\bibitem [{\citenamefont {Bruschi}(2019)}]{bruschi_time_2019}%
  \BibitemOpen
  \bibfield  {author} {\bibinfo {author} {\bibfnamefont {D.~E.}\ \bibnamefont {Bruschi}},\ }\bibfield  {title} {\bibinfo {title} {Time evolution of coupled multimode and multiresonator optomechanical systems},\ }\href {https://doi.org/10.1063/1.5106409} {\bibfield  {journal} {\bibinfo  {journal} {Journal of Mathematical Physics}\ }\textbf {\bibinfo {volume} {60}},\ \bibinfo {pages} {062105} (\bibinfo {year} {2019})}\BibitemShut {NoStop}%
\bibitem [{\citenamefont {Bose}\ \emph {et~al.}(1997)\citenamefont {Bose}, \citenamefont {Jacobs},\ and\ \citenamefont {Knight}}]{bose_preparation_1997}%
  \BibitemOpen
  \bibfield  {author} {\bibinfo {author} {\bibfnamefont {S.}~\bibnamefont {Bose}}, \bibinfo {author} {\bibfnamefont {K.}~\bibnamefont {Jacobs}},\ and\ \bibinfo {author} {\bibfnamefont {P.~L.}\ \bibnamefont {Knight}},\ }\bibfield  {title} {\bibinfo {title} {Preparation of nonclassical states in cavities with a moving mirror},\ }\href {https://doi.org/10.1103/PhysRevA.56.4175} {\bibfield  {journal} {\bibinfo  {journal} {Physical Review A}\ }\textbf {\bibinfo {volume} {56}},\ \bibinfo {pages} {4175} (\bibinfo {year} {1997})}\BibitemShut {NoStop}%
\bibitem [{\citenamefont {Blencowe}(2013)}]{blencowe_effective_2013}%
  \BibitemOpen
  \bibfield  {author} {\bibinfo {author} {\bibfnamefont {M.~P.}\ \bibnamefont {Blencowe}},\ }\bibfield  {title} {\bibinfo {title} {Effective {Field} {Theory} {Approach} to {Gravitationally} {Induced} {Decoherence}},\ }\href {https://doi.org/10.1103/PhysRevLett.111.021302} {\bibfield  {journal} {\bibinfo  {journal} {Physical Review Letters}\ }\textbf {\bibinfo {volume} {111}},\ \bibinfo {pages} {021302} (\bibinfo {year} {2013})}\BibitemShut {NoStop}%
\bibitem [{\citenamefont {Xu}\ and\ \citenamefont {Blencowe}(2022)}]{xu_zero-dimensional_2022}%
  \BibitemOpen
  \bibfield  {author} {\bibinfo {author} {\bibfnamefont {Q.}~\bibnamefont {Xu}}\ and\ \bibinfo {author} {\bibfnamefont {M.~P.}\ \bibnamefont {Blencowe}},\ }\bibfield  {title} {\bibinfo {title} {Zero-dimensional models for gravitational and scalar {QED} decoherence},\ }\href {https://doi.org/10.1088/1367-2630/aca427} {\bibfield  {journal} {\bibinfo  {journal} {New Journal of Physics}\ }\textbf {\bibinfo {volume} {24}},\ \bibinfo {pages} {113048} (\bibinfo {year} {2022})},\ \bibinfo {note} {arXiv:2005.02554 [gr-qc, physics:quant-ph]}\BibitemShut {NoStop}%
\bibitem [{\citenamefont {Barros~e Sá}\ and\ \citenamefont {Gomes}(2021)}]{barros_e_sa_quantum_2021}%
  \BibitemOpen
  \bibfield  {author} {\bibinfo {author} {\bibfnamefont {N.}~\bibnamefont {Barros~e Sá}}\ and\ \bibinfo {author} {\bibfnamefont {C.}~\bibnamefont {Gomes}},\ }\bibfield  {title} {\bibinfo {title} {From quantum field theory to quantum mechanics},\ }\href {https://doi.org/10.1140/epjc/s10052-021-09742-0} {\bibfield  {journal} {\bibinfo  {journal} {The European Physical Journal C}\ }\textbf {\bibinfo {volume} {81}},\ \bibinfo {pages} {931} (\bibinfo {year} {2021})}\BibitemShut {NoStop}%
\end{thebibliography}%
\end{document}